\documentclass[aps,prd,reprint,superscriptaddress,nofootinbib]{revtex4-1}

\usepackage{amsmath}
\usepackage[dvipsnames]{xcolor}
\usepackage{graphicx}
\usepackage{hyperref}
\usepackage{mathtools}
\usepackage{comment}
\usepackage{cancel}
\usepackage{physics}
\usepackage{mathtools}
\usepackage{adjustbox}
\usepackage{multirow,array,booktabs}
\usepackage[shortlabels]{enumitem}
\usepackage{enumitem} 

\hypersetup{
 bookmarksnumbered=true,
 colorlinks=true,
 linkcolor=[rgb]{0.098,0.098,0.439},
 citecolor=[rgb]{0.098,0.098,0.439},
 urlcolor=[rgb]{0.098,0.098,0.439}
  }

\begin{document}

\title{Refined approaches in second leptogenesis for the baryon-lepton asymmetry discrepancy}

\author{YeolLin ChoeJo}
\email{particlephysics@kaist.ac.kr}
\affiliation{Department of Physics, Korea Advanced Institute of Science and Technology, Daejeon 34141, Korea}
\author{Kazuki Enomoto}
\email{k\_enomoto@phys.ntu.edu.tw}
\affiliation{Department of Physics, Korea Advanced Institute of Science and Technology, Daejeon 34141, Korea}\author{Yechan Kim}
\email{cj7801@kaist.ac.kr}
\affiliation{Department of Physics, Korea Advanced Institute of Science and Technology, Daejeon 34141, Korea}\author{Hye-Sung Lee}
\email{hyesung.lee@kaist.ac.kr}
\affiliation{Department of Physics, Korea Advanced Institute of Science and Technology, Daejeon 34141, Korea}

\begin{abstract}
The temperature-dependent mass of the heavy neutrino can lead to the second leptogenesis occurring below the electroweak scale, potentially explaining the large discrepancy between baryon and lepton asymmetries. We investigate this scenario further, exploring the intricate interplay of the weak interaction processes within this framework.
It includes notable shifts in the dominant decay channels of heavy neutrinos around the electroweak symmetry breaking, along with the resonance behavior of the scattering processes near the $W/Z$ mass. The $CP$ asymmetry can also vary over cosmic history due to the temperature-dependent mass, allowing the $B-L$ asymmetry generation to be amplified in the late epoch.
These findings elucidate how such alterations in the dynamics of second leptogenesis contribute to addressing the observed discrepancies in baryon-lepton asymmetry.
\end{abstract}

\maketitle

\section{Introduction}
\label{Intro}

Our Universe exhibits a nonzero matter-antimatter asymmetry, and the mechanism behind this asymmetry remains a crucial open question in particle physics. One of the reasonable new physics scenarios to explain it is leptogenesis \cite{Fukugita:1986hr}, where a nonzero lepton asymmetry is generated by the out-of-equilibrium decay of heavy Majorana neutrinos with lepton number violation. The created lepton asymmetry is then partially converted into that in the baryon number via the sphaleron transition in equilibrium at finite temperatures~\cite{Manton:1983nd, Klinkhamer:1984di, Kuzmin:1985mm}, which leads to the baryon asymmetry comparable to the lepton asymmetry.  
This scenario provides an explanation for the observed baryon asymmetry, supplemented by the generation of the tiny Majorana masses of the active neutrinos by the type-I seesaw mechanism~\cite{Minkowski:1977sc, Yanagida:1979as, Gell-Mann:1979vob, Mohapatra:1979ia, Schechter:1980gr}.

The amount of baryon asymmetry of the Universe (BAU) is quantified by the baryon-to-photon ratio, $\eta_B \equiv \frac{n_B - n_{\bar{B}}}{n_\gamma} \simeq \mathcal{O}(10^{-10})$, where $n_B$, $n_{\bar{B}}$, and $n_\gamma$ are the number densities of baryons, antibaryons, and photons, respectively. 
The value of it is determined by the measurements of the light elements' abundances in the Universe~\cite{Fields:2019pfx} and the observation of the cosmic microwave background (CMB) \cite{Planck:2018vyg}.  On the other hand, the latest ${}^4 \mathrm{He}$ abundance observation at the EMPRESS experiment~\cite{Matsumoto:2022tlr} suggests that the lepton asymmetry of the Universe (LAU) is significantly larger than the BAU. 
They reported the value of the degeneracy parameter of the electron neutrino as $\xi_{\nu_e} = 0.05^{+0.03}_{-0.02}$. 
It relates to the lepton-to-photon ratio $\eta_L$ as
\begin{equation}
\label{eq: EtaL}
\eta_L \equiv \frac{n_L - n_{\bar{L}}}{n_\gamma} \simeq \sum_\ell \frac{g_\ell \pi^2}{12 \zeta(3)} \left( \frac{T_\ell}{T_\gamma} \right)^3 \xi_\ell,
\end{equation}
where $g_\ell$ and $T_\ell$ represent the degree of freedom and temperature of lepton $\ell$, i.e., charged leptons and neutrinos. 
By using appropriate assumptions, the EMPRESS results leads to $\eta_L = \mathcal{O}(10^{-2})$ with an uncertainty of $2.5~\sigma$ level~\cite{Kawasaki:2022hvx, Burns:2022hkq, Borah:2022uos, Escudero:2022okz, ChoeJo:2023cnx}. 

Although the suggested large LAU is still an anomaly of the $2.5 \sigma$ level, the difference between the BAU and LAU is considerably large, $\eta_L \simeq 10^8 \eta_B$. Since the sphaleron transition in equilibrium equates the orders of these two asymmetries~\cite{Kuzmin:1985mm}, such a large discrepancy calls for modifications to the traditional scenario of leptogenesis. See Refs.~\cite{
Kawasaki:2022hvx, Burns:2022hkq, Borah:2022uos, Escudero:2022okz, Takahashi:2022cpn, Kasuya:2022cko, Domcke:2022uue, Deng:2023twb, Gao:2023djs, ChoeJo:2023cnx, Jiang:2024zrb} for further discussions on the measurements of lepton asymmetry and efforts toward its explanation.

To explain the discrepancy, a new scenario was proposed involving twofold leptogenesis, enabled by the time-varying masses of heavy neutrinos due to their coupling with scalar wave dark matter (wave DM) $\phi$ \cite{ChoeJo:2023cnx}. 
In the early Universe, the masses of heavy neutrinos were initially constant and much larger than their bare Majorana masses due to the coupling with $\phi$. 
At a certain temperature $T_\ast$, $\phi$ started to oscillate, whose amplitude decreased with temperature cooling due to cosmic expansion.
Then, the size of the heavy neutrino masses also began to decrease. 
The change of the mass ended when the temperature-dependent mass was comparable to the bare mass. 
As a result, there are two ``bendings'' in the curve of the heavy neutrino mass throughout the history of the Universe. In Refs.~\cite{Bhandari:2023wit, Zhao:2024uid}, leptogenesis with such a temperature-dependent mass has also been discussed. 

\begin{figure}[tb]
\centering
\includegraphics[width=0.49\textwidth]{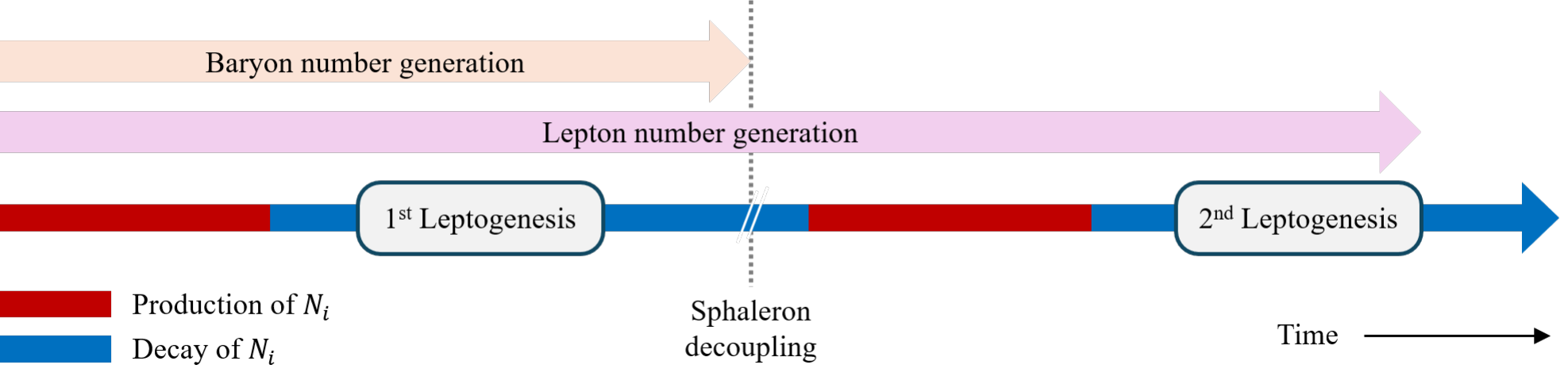}
\caption{
The schematic diagram for the decoupling of baryon and lepton asymmetry in the twofold leptogenesis scenario. If sphaleron decoupling occurs between the first and second phases of leptogenesis, the baryon asymmetry is fixed earlier than the lepton asymmetry, while the lepton number can still be altered through the second leptogenesis.
}
\label{SchematicPlot}
\end{figure}

The temperature-dependent Majorana mass enables two distinct eras where heavy neutrinos are nonrelativistic in the early Universe. 
In such eras, the production of heavy neutrinos was extremely suppressed. 
The decay occurred dominantly, and the nonzero lepton number was created if the decay was nonequilibrium and included $CP$ violation. 
Therefore, in suitable parameter regions, we would expect that leptogenesis occurred twice in the early Universe. 
In addition, it is possible to consider a setup where the first leptogenesis occurred before the electroweak symmetry breaking (EWSB), and the second one occurred after the EWSB. 
This decouples the size of the BAU and LAU because the sphaleron transition decoupled after EWSB, and the lepton asymmetry generated at the second leptogenesis was not converted into baryon asymmetry. 
It provides a plausible scenario to explain the large discrepancy between the BAU and LAU.
The schematic diagram for this scenario is illustrated in Fig.~\ref{SchematicPlot}.

Such an additional production of lepton number after EWSB can also be achieved via leptogenesis through neutrino oscillations~\cite{Akhmedov:1998qx} in minimal setups without introducing new scalar fields, as discussed in Refs.~\cite{Shaposhnikov:2008pf, Canetti:2012vf, Canetti:2012kh, Ghiglieri:2019kbw, Ghiglieri:2020ulj, Eijima:2020shs}. This process plays a crucial role in the neutrino minimal standard model ($\nu$MSM) \cite{Asaka:2005an, Asaka:2005pn}, where it nonthermally produces the lightest sterile neutrino \cite{Dodelson:1993je, Shi:1998km} that acts as dark matter in the model.

In the $\nu$MSM, the lightest sterile neutrino has a much lighter mass than the other two and possesses a tiny coupling with the standard model (SM) particles, remaining decoupled at all times. In contrast, in our scenario, all three sterile neutrinos have similar masses, and their Yukawa couplings are as large as those expected from the seesaw mechanism because we introduce the wave dark matter $\phi$. The size of the lepton asymmetry is independent of the dark matter relic abundance in our scenario, whereas they are correlated in the $\nu$MSM. Therefore, our work presents a new approach in models with sterile neutrinos to address tiny neutrino masses, DM, BAU, and the large lepton asymmetry.

Furthermore, the temperature-dependent mass of the sterile neutrinos can change the sign of the lepton asymmetry before and after sphaleron decoupling. This is essential to explain both the positive baryon asymmetry, which is generated from a negative lepton asymmetry, and the positive lepton asymmetry.

In this paper, we perform a detailed analysis of this new leptogenesis scenario, with accurate consideration of the electroweak interaction modes of heavy neutrinos $N_i$ after the EWSB. We discuss how it changes the nature of the second leptogenesis.

This paper is organized as follows. In Sec.~\ref{Model}, we present the Lagrangian of the model and discuss the behavior of the temperature-dependent mass of the heavy neutrinos. In Sec.~\ref{DecayScattering}, we explore changes in the decay and scattering processes of $N_i$ due to its mass variation and mixing with the SM neutrinos. 
In Sec.~\ref{DME}, we show the density matrix equations and $CP$ asymmetry adopting the change of the decay and scattering channels.
In Sec.~\ref{Estimation}, we perform the numerical evaluation of the baryon and lepton asymmetry. Finally, we summarize our findings and discuss the outlook in Sec.~\ref{Summary}. 
Detailed formulas for the scattering cross sections are provided in Appendix~\ref{Appendix}.
In Appendix~\ref{sec: app_CPasym}, a general discussion about the $CP$ asymmetry in the decay of the heavy neutrinos is presented. 

\section{Model for the second leptogenesis}
\label{Model}

We add three gauge singlet right-handed neutrinos $N_R^i$ ($i=1,2,3$) and a scalar $\phi$ into the SM. We assume that $\phi$ is wave dark matter with the mass $10^{-22}~\mathrm{eV} < m_\phi < 30~\mathrm{eV}$~\cite{Hui:2021tkt}, where the lower bound is given by the Lyman-$\alpha$ forest data, and the upper bound comes from the consistency of using the classical field to describe $\phi$. 
The energy density of the coherent oscillation of $\phi$ behaves like matter during cosmic expansion and constitutes the current dark matter density. 
The terms in the Lagrangian relevant to the following discussion are given by~\cite{Krnjaic:2017zlz, Dev:2022bae, ChoeJo:2023ffp}
\begin{equation}
\label{interaction}
\mathcal{L} = - \frac{1}{2} (M_{0i} + g_i \phi) \overline{N^{ic}_R} N^i_R - y_{\ell i} \overline{L_\ell} \Phi^c N_R^i + \mathrm{H.c.}
\end{equation}
where $M_{0i}$ and $g_i$ are the bare Majorana masses and coupling constants of the heavy neutrinos, respectively. 
The Higgs doublet and the lepton doublet are denoted by $\Phi$ and $L_\ell$, respectively, and $\Phi^c$ is defined by $\Phi^c = i \sigma_2 \Phi^\ast$. 

After the EWSB, the Higgs field acquires the vacuum expectation value (VEV) $v \simeq 246~\mathrm{GeV}$, which generates the Dirac mass $(m_D)_{\ell i} = y_{\ell i} v /\sqrt{2}$ for neutrinos. 
Then, the left-handed neutrinos $\nu_L^\ell$ in $L_\ell$ are expressed by the mass eigenstate Majorana fields $\nu_a$ and $N_i$ ($a, i=1,2,3$) as
\begin{align}
\nu_L^\ell & = U_{\ell a} P_L \nu_a + R_{\ell i} P_L N_i,
\label{ActiveSterile}
\end{align} 
where $P_L$ is the left-handed chirality projection operator, $U$ is the Pontecorvo-Maki-Nakagawa-Sakata (PMNS) matrix~\cite{Pontecorvo:1957cp, Maki:1962mu}, and $R$ is the active-sterile mixing matrix. 
At the leading order, $R$ is given by 
\begin{equation}
R = m_D M_N^{-1}, 
\end{equation}
where $M_N$ is the diagonal mass matrix of the heavy neutrinos. 

Assuming spatial homogeneity, the configuration of the classical field $\phi(t)$ is determined by the following equation of motion,
\begin{equation}
\ddot{\phi} + 3H\dot{\phi} + m_{\phi}^2 \phi = 0, 
\end{equation}
where the dot means the time derivative and $H$ is the Hubble parameter. 
In the early Universe, $H$ is much larger than $m_\phi$, and $\phi(t)$ was a constant due to the large friction term. In this paper, we simply assume that the initial configuration $\phi_0$ is nonzero and do not discuss its origin. 
Since $H$ decreases as the Universe expands, $\phi$ begins to oscillate at a specific temperature $T=T_*$, which satisfies $H(T_*) = m_\phi$. 
At lower temperatures, $\phi$ oscillates as $\phi \propto \cos(m_\phi t)$, and its energy density $\rho_\phi = \frac{1}{2} \dot{\phi}^2 + \frac{1}{2} m_\phi^2 \phi^2$ scales as a matter component with respect to the cosmic expansion. 
The amplitude of the oscillation $\phi_\mathrm{amp}$ scales as $\phi_\mathrm{amp} \propto \sqrt{\rho} \propto T^{3/2}$.

A nonzero configuration of $\phi$ can change the mass of $N_i$ via the couplings in Eq.~(\ref{interaction}). 
At $T > T_\ast$, $N_i$ has a constant mass $M_{\ast i}$ determined by the initial configuration of $\phi$, $M_{\ast i} = M_{0i} + g_i \phi_0$. 
In the following, we consider the case that $M_{\ast i}$ is much larger than the bare mass $M_{0i}$.
At $T < T_\ast$, the oscillation of $\phi$ induces the time-(temperature-)dependent mass of the heavy neutrinos, $M_i(t) = M_{0i} + g_i \phi(t)$.  

When $\phi(t)$ becomes negative, $N_i$ formally acquires a negative mass if the amplitude is sufficiently large, i.e., $M_i \simeq g_i \,\phi_{\mathrm{amp}}\,\cos(m_{\phi} t)$. However, the overall sign of the mass is physically irrelevant because one can always rephase the right-handed fields to flip the sign. Therefore, in analyzing the effect of the oscillating $\phi$, we focus on its magnitude rather than simply on $g_i\,\phi(t)$.

Moreover, it is convenient to simplify the problem by averaging over one oscillation period. The amplitude $\phi_{\mathrm{amp}}$ changes on a time scale $t_{\mathrm{amp}} \sim 1/H$, as indicated by the equation of motion $d\phi_{\mathrm{amp}}/dt = \phi_{\mathrm{amp}}\,H$. In contrast, the oscillation itself has a period $t_{\mathrm{osc}} \sim 1/m_{\phi}$. Once the oscillation begins, we have $H < m_{\phi}$, so the field oscillates many times before its amplitude changes significantly. Consequently, it is a good approximation to use the time-averaged oscillation in subsequent calculations.

Concretely, for the mass variation term $g_i\,\phi(t)$, one may use $\langle |\,g_i\,\phi(t)| \rangle$ or $\langle g_i^2\,\phi^2(t) \rangle^{1/2}$ as the effective mass scale, where $\phi(t) = \phi_{\mathrm{amp}}\cos(m_{\phi}t)$ and $\langle \cdots \rangle$ denotes a time average.  Here, we adopt the squared average, $\langle g_i^2 \,\phi^2(t)\rangle^{1/2}$.\footnote{
Other averaging choices, such as $\langle |M_i(t)|\rangle$ or $\langle M_i^2(t)\rangle^{1/2}$, yield similar results for both high and low temperatures. 
They do differ somewhat from our choice when $M_i \simeq M_{0i}$ [i.e., $T \simeq T_{N_i}$ in Eq.~(\ref{Mcase})], but this discrepancy does not significantly affect leptogenesis in our scenario.}

As a result, wave dark matter leads to the following temperature-dependent mass of $N_i$ for $T<T_\ast$; 
\begin{align}
M_i & \simeq M_{0i} + g_i |\phi| \left<\cos^2 (m_\phi t)  \right>^{1/2} 
\\
& = M_{0i} + g_i \frac{ \sqrt{\rho_\mathrm{DM}^{} } }{ m_\phi } \left( \frac{ T }{ T_0 } \right)^{3/2}, 
\label{eq: temperature mass}
\end{align}
where $\rho_\mathrm{DM}^{}$ is the dark matter energy density, and $T_0$ is the temperature of the current Universe. 
The mass behavior of $N_i$ in the early Universe is summarized by~\cite{ChoeJo:2023cnx}
\begin{equation}
\label{Mcase}
M_i(T) \simeq
\begin{cases}
M_{\ast i}  & T > T_\ast, \\
M_{0i} + M_{*i} \left(\frac{T}{T_*} \right)^{3/2} & T_\ast > T > T_{N_i}, \\
M_{0i} & T_{N_i} > T,
\end{cases}
\end{equation}
where $T_{N_i}$ is defined by $M_i(T_{N_i}) = 2M_{0i}$, below which the contribution from the wave dark matter coupling is negligibly small compared to the bare mass. 

To justify the use of the time-average approximation of the mass in studying leptogenesis, 
another condition needs to be satisfied; the timescale of the $N_i$ decays has to be much longer than that of the oscillation $t_\mathrm{osc} \sim 1/m_\phi$, i.e., the decay rate has to be smaller than $m_\phi$. 
We have checked that this condition is satisfied 
at the parameter point investigated in Sec.~\ref{Estimation}.

Before closing this section, we briefly discuss various constraints on the model parameters. 
First, the mass of $N_i$ and the size of the Yukawa coupling $y_{\ell i}^{}$ are constrained by the big bang nucleosynthesis (BBN) and $N_\mathrm{eff}$~\cite{Vincent:2014rja, Bolton:2019pcu}.  
To avoid disrupting their predictions due to the decay after the neutrino decoupling, $M_{0i}$ has to be larger than roughly $1~\mathrm{GeV}$ when the size of the Yukawa coupling satisfies the canonical seesaw relation~\cite{Boyarsky:2009ix, Ruchayskiy:2012si}. 
To avoid this constraint, we consider $M_{0i} \simeq 10~\mathrm{GeV}$ in the following and assume the decay product of $N_i$ is thermalized before the BBN.

To make $\phi$ stable, the decay $\phi \to N_i N_i$ has to be prohibited, or its lifetime has to be much larger than the cosmic age. 
In the following, we assume $M_i \geq M_{0i} \simeq 10~\mathrm{GeV}$ and $m_\phi \simeq 10^{-3}~\mathrm{eV}$. 
Thus, the decay does not occur at zero temperature. 
Even at finite temperatures, the thermal mass of $\phi$ induced by the thermalized background $N_i$ does not exceed the criterion of the decay because we consider very small $g_i$, and $N_i$ is thermalized only around $T \simeq M_{0i} \simeq 10~\mathrm{GeV}$. 
Thus, the decay was also prohibited in the early Universe. 

To keep the coherent oscillation of $\phi$ in the entire cosmic history, the thermalization of $\phi$ via the scattering $\phi N_i \rightarrow \phi N_i$ has to be avoided~\cite{Dev:2022bae}. 
In addition, for $\phi$ to behave as wave dark matter, the one-loop induced quartic coupling of $\phi$ should be smaller than its mass term at the matter-radiation equality~\cite{Dev:2022bae}. 
These constrain the model parameters. 
For instance, at $T_\ast = 1$ TeV, the values of $M_{\ast i} \gtrsim 1.6 \times 10^7$ GeV are excluded due to the dominant quartic coupling. Furthermore, $M_{\ast i} \gtrsim 4.5 \times 10^7$ GeV $(M_0 / 10 \; \mathrm{GeV})^{1/4}$ are excluded due to the thermalization of $\phi$. See Ref.~\cite{ChoeJo:2023cnx} for more details. 

There are other constraints from the experiments such as Majoron detection~\cite{GERDA:2022ffe}, neutrino free-streaming~\cite{Huang:2021kam}, and neutrino oscillations in the Sun~\cite{Berlin:2016woy}. 
However, they are not strong constraints in the parameter regions discussed below. 
Thus, we present no further discussion here.

\section{Decay and scattering channels of heavy neutrinos in a temperature-dependent mass scenario}
\label{DecayScattering}

In this section, we discuss the decay and scattering processes of the heavy neutrinos.
Depending on the temperature and mass, different processes may dominate the production of $N_i$ in the early Universe.

\subsection{Decay channels}

Here, we show the formulas for the decay rates of $N_i$. 
Since we discuss them at the tree level, the decay to particles and that to antiparticles have the same decay rate. 
Thus, we show only the formula for the decay violating the lepton number by $\Delta L = 1$. 
To evaluate the total decay rate, we need to double each formula to include $\Delta L = -1$ modes. 
In addition, all fermion flavors are summed in the formulas.

Before the EWSB, the Majorana fermions $N_i = N_R^i + (N_R^i)^c$ are the massive mode with the temperature mass $M_i(T)$, and the other particles are massless.\footnote{We neglect the thermal mass in the following discussion.}
The dominant decay modes are $N_i \rightarrow L_\ell \Phi$. The decay rate is given by
\begin{equation}
\label{eq: decay_to_doublets}
\Gamma (N_i \to L \Phi) = \frac{ (y^\dagger y)_{ii} M_i }{16\pi}, 
\end{equation}
where we took the sums about the $SU(2)$ doublet. 

After the EWSB, the Higgs field obtains the VEV. 
The Dirac mass terms for neutrinos are generated, and the neutrino mixing shown in Eq.~(\ref{ActiveSterile}) occurs. 
Through the mixing, $N_i$ interacts with the electroweak gauge bosons $W^\pm/Z$ and the Higgs boson $h$. At the leading order of $M_i^{-1}$, the interaction terms are given by
\begin{align}
\label{eq: N_interactions_afterEWSB}
\mathcal{L} \simeq & - \frac{g_W^{}}{\sqrt{2}} (m_D M_{N}^{-1})_{a i} \bar{\ell}_a \gamma^\mu P_L N^i W_{\mu}^-  \nonumber
\\
& - \frac{g_Z^{}}{2} (m_D M_{N}^{-1})_{ai} \overline{\nu_a} \gamma^\mu P_L N^i Z_{\mu}  \nonumber \\[5pt]
& - \frac{ 1}{ v } (m_D)_{ai} h \overline{\nu_a} P_R N_i + \mathrm{H.c.},
\end{align}
where $g_W^{}$ is the charged weak coupling, and $g_Z^{}$ is the neutral weak coupling.
Since the couplings of the gauge interactions include $M_N^{-1}$, the strength of the gauge interaction also depends on the temperature. 
We do not show the PMNS matrix in Eq.~(\ref{eq: N_interactions_afterEWSB}) because we neglect the active neutrino masses in the following discussions.\footnote{It does not mean that we approximate $m_D \simeq 0$. We neglect the kinematic effects of the active neutrino mass. Then, the PMNS matrix does not appear in the flavor-summed formulas.} 
The active neutrinos $\nu$ are treated as Weyl fermions, not Majorana fermions. 
Thus, we distinguish $\nu$ and the antineutrinos $\bar{\nu}$.

If $N_i$ is heavier than $Z$, $W^\pm$, and $h$, then the corresponding two-body decays are kinematically allowed. 
The decay rates are given by
\begin{align}
\label{Znu}
& \Gamma(N_i \rightarrow Z \nu) = \frac{(y^\dagger y)_{i i}  M_i}{64 \pi  } \left(1 - 3 \frac{m_Z^4}{M_{i}^4} + 2 \frac{m_Z^6}{M_{i}^6} \right) , \\[5pt]
\label{Well}
& \Gamma(N_i \rightarrow W^+ \ell)  = \frac{(y^\dagger y)_{i i} M_i}{64 \pi } \left( 1 - 3 \frac{m_W^4}{M_{i}^4} + 2 \frac{m_W^6}{M_{i}^6} \right) , \\[5pt]
\label{Hnu}
& \Gamma(N_i \rightarrow h \nu)  = \frac{ (y^\dagger y)_{i i} M_{i} }{64 \pi }  \left( 1 - \frac{m_h^2}{ M^2_{i} } \right)^2 ,
\end{align}
where $m_Z$, $m_W$, and $m_h$ are the masses of the $Z$, $W$, and Higgs bosons, respectively. In the above expressions, the lepton masses are neglected.

When the above two-body decays are not allowed, the dominant decay processes are three-body decays via off-shell $Z$, $W^\pm$, and $h$ such as $N \rightarrow \nu\nu\bar{\nu}$, $N \rightarrow \nu\ell\bar{\ell}$, $N \rightarrow \nu q\bar{q}$, and $N \rightarrow \ell q\bar{q}'$ \cite{johnson1997extending, gorbunov2007find, Gelmini:2020ekg}. 
Since we consider $N_i$ with a mass larger than $M_{0i} \simeq 10~\mathrm{GeV}$, the energy scale of the decay exceeds the QCD scale. 
Thus, the final states include free quarks, not hadronic states. 

From Eq.~(\ref{eq: N_interactions_afterEWSB}), the ratio of the strength of the interactions with $Z$, $W^\pm$, $h$ are roughly given by 
\begin{align}
g_{W^\pm N_i}^{}: g_{Z N_i}^{}: g_{h N_i}^{} 
& \simeq \frac{ g_W }{ \sqrt{2} M_i }: \frac{ g_Z }{ 2M_i }: \frac{ 1 }{ v } \nonumber
\\
& \simeq \frac{\sqrt{2} m_W^{} }{ M_i}: \frac{ m_Z^{} }{M_i}: 1 .
\end{align}
Since we consider the case that two-body decays are not allowed, the gauge bosons give a larger contribution than the Higgs boson. 
Thus, in the following, we consider only the three-body decays via the off-shell gauge bosons. 

In evaluating the decay rates, we use the approximation of the contact Fermi couplings for the internal gauge bosons. In addition, we neglect all the fermion masses. We do not consider decays involving the top quarks because they are assumed to be decoupled simultaneously after the EWSB for simplicity. 
Then, the decay rates are given by
\begin{align}
\label{3nu}
\Gamma(N_i \rightarrow \nu \nu \bar{\nu}) = & \frac{G_F M_i^3}{192 \sqrt{2}\pi^3} (y^\dagger y)_{ii} , \\[5pt]
\Gamma(N_i \rightarrow \nu \ell \bar{\ell}) = & \frac{G_F M_i^3}{768\sqrt{2}\pi^3}  (y^\dagger y)_{ii} 
\\
& \quad \times \left(11-4\sin^2\theta_W^{} + 24\sin^4\theta_W^{} \right), \nonumber
\\
\Gamma(N_i \rightarrow \nu u \bar{u}) = & \frac{3 G_F M_i^3}{384 \sqrt{2}\pi^3}  (y^\dagger y)_{ii} 
\\
& \quad \times
\left(1-\frac{8}{3}\sin^2\theta_W^{}+\frac{32}{9}\sin^4\theta_W^{} \right), \nonumber \\
\Gamma(N_i \rightarrow \nu d \bar{d}) = & \frac{9 G_F M_i^3}{768\sqrt{2}\pi^3}  (y^\dagger y)_{ii} 
\\
& \quad \times
\left(1-\frac{4}{3}\sin^2\theta_W^{} +\frac{8}{9}\sin^4\theta_W^{} \right), \nonumber \\[5pt]
\label{lqq}
\Gamma(N_i \rightarrow \ell u \bar{d}) = & \sum_{\alpha, \beta} \frac{3G_F M_i^3}{192 \sqrt{2} \pi^3}  (y^\dagger y)_{ii} \abs{V_{\alpha\beta}}^2, 
\end{align}
where the $u$ and $d$ represent up-type and down-type quarks, respectively, $V_{\alpha\beta}$ denotes the Cabbibo-Kobayashi-Maskawa matrix element~\cite{Cabibbo:1963yz, Kobayashi:1973fv}, and $\theta_W^{}$ is the weak mixing angle.

The phase space suppression due to the fermion masses is neglected in the above formulas~\cite{gorbunov2007find, johnson1997extending}.
Since we consider $M_{0i} \simeq 10~\mathrm{GeV}$, 
it would be important for the decays including the bottom quarks ($b$), charm quarks ($c$), and tau leptons ($\tau$) when $M_i(T) \simeq M_{0i}$. 
However, the massless approximation is still a good approximation for other decay channels, including only the light fermions, and the number of such decay modes is much larger than those including $b$, $c$, and $\tau$. 
Thus, the order of the total decay rate does not change, and we expect that phase space suppression has little impact on evaluating leptogenesis. 

Finally, we comment on the effects of the background plasma. Below the electroweak scale, there is a matter effect on the active-sterile mixing. However, it is negligible within our relevant mass range \cite{Gelmini:2020ekg}, and thus, we do not consider it in this paper.

\subsection{Scattering channels}
\label{sec: scattering}

Here, we discuss the scattering processes involving one $N_i$. 
We only consider the processes with $\Delta L = -1$. It can also be used to evaluate the effect of the inverse $\Delta L = +1$ processes in leptogenesis, as discussed in Sec.~\ref{Estimation}. 

Before the EWSB, the dominant scattering processes is $NL \leftrightarrow Q_3 t$ and $N Q_3 \to L t$ via $\Phi$ exchange, where $Q_3$ is the third-generation quark doublet and $t$ is the right-handed top quark. 
The scattering rate at finite temperatures is roughly given by 
\begin{equation}
\Gamma_\mathrm{prod} \simeq \frac{ y_t^2 }{ 4 \pi } (y^\dagger y)_{ii} T, 
\end{equation}
where $y_t \simeq 1.16$ is the top quark Yukawa coupling~\cite{CMS:2020djy}.
As shown in Fig.~\ref{RatePlot} in Sec.~\ref{Estimation}, the effect of these scattering is smaller than the decay in relevant parameter regions. Thus, we employ this approximated formula and do not use their exact cross section formula to evaluate the $N_i$ production via these processes.

After the EWSB, $N_i$ has the gauge interaction. 
Thus, the $2 \to 2$ scattering processes via $Z$ and/or $W^\pm$ are allowed. 
The relevant $\Delta L = -1$ scattering processes are $N\nu \rightarrow \ell\bar{\ell}/\nu\bar{\nu}/q\bar{q}$, $N\bar{\nu} \rightarrow 2\bar{\nu}$, $N\ell \rightarrow \bar{\nu} \ell/\bar{u}d$, $N\bar{\ell} \rightarrow \bar{\nu} \bar{\ell}$, $Nq (\bar{q}) \rightarrow \bar{\nu}q (\bar{q})$, $N \bar{u} \to \bar{\ell} d$, and $N\bar{d} \to \bar{\ell}\bar{u}$. $\Delta L = +1$ processes are given by their $CP$ conjugate processes. 
Since these give a significant contribution to leptogenesis, we evaluate the $N_i$ production via these scatterings by using their cross section formula and an exact integral expression as discussed in the following section.
The formulas for the scattering cross section of each process are shown in Appendix~\ref{Appendix}.

\section{Density matrix equations and $CP$ asymmetry in the decays}
\label{DME}

Now, we present the analysis for the second leptogenesis scenario along the cosmic timeline, incorporating these temperature-dependent decay and scattering rates of $N_i$. 
We consider the amount of $N_i$ and $B-L$ asymmetry in a portion of comoving volume defined in Ref.~\cite{Buchmuller:2002rq} which are denoted by $N_{N_i}^{}$ and $N_{\alpha \beta}$ ($\alpha, \beta = e$, $\mu$, $\tau$) in the following. 
The diagonal term $N_{\alpha \alpha}$ represents the number of $B/3 - L_\alpha$ in the comoving volume, where $L_\alpha$ is the lepton number with the flavor $\alpha$. 
The off-diagonal terms represent the coherence among the lepton flavors. 
The evolution of $N_{N_i}$ and $N_{\alpha\beta}$ is described by the following density matrix equations~\cite{Abada:2006fw, DeSimone:2006nrs, Blanchet:2006ch, Blanchet:2011xq, Moffat:2018wke, Granelli:2021fyc, Granelli:2023egb}
\begin{align}
\frac{ \mathrm{d} N_{N_i} }{ \mathrm{d} z } = & - (D_i + S_i) (N_{N_i} - N_{N_i}^\mathrm{eq} ),
\label{Neqnres}
\\
\frac{ \mathrm{d} N_{\alpha \beta} }{ \mathrm{d} z } = & -\sum_i \left[ \varepsilon^{(i)}_{\alpha \beta} D_i (N_{N_i} - N_{N_i}^\mathrm{eq} )
+ \frac{1}{2} W_i \{ P_i, N \}_{\alpha \beta} \right] \nonumber \\
& - \frac{\Gamma_{\tau}}{Hz} [I_{\tau}, [I_{\tau}, N]]_{\alpha\beta}
- \frac{\Gamma_{\mu}}{Hz} [I_{\mu}, [I_{\mu}, N]]_{\alpha\beta},
\label{NBLeqnres}
\end{align}
where $z \equiv M_{01}/T$, $N_{N_i}^\mathrm{eq}$ is $N_{N_i}$ in thermal equilibrium, $D_i$ and $S_i$ express effects of the decays and scatterings, respectively, $W_i$ is the washout term evaluated by $D_i$ and $S_i$ as shown below, $\Gamma_\tau$ and $\Gamma_\mu$ are effects of the tau and muon Yukawa interactions, respectively, and $\varepsilon_{\alpha \beta}^{(i)}$ is the $CP$ asymmetry in the decay of $N_i$.\footnote{In our convention, $\varepsilon_{\alpha \beta}^{(i)}$ has the opposite sign to that employed in Refs.~\cite{Abada:2006fw, DeSimone:2006nrs, Blanchet:2006ch, Blanchet:2011xq,  Moffat:2018wke, Granelli:2021fyc, Granelli:2023egb}. See Appendix~\ref{sec: app_CPasym}.} 
In the following, we explain each quantity in the equations in detail. 

$D_i$ is given by  
\begin{equation}
    D_i = \frac{ K_1 (z) }{ K_2 (z) } \frac{ \Gamma_i }{ Hz}, 
\end{equation}
where $\Gamma_i$ is the total decay rate of $N_i$, and $K_n$ is the $n$th modified Bessel function of the second kind. 
The prefactor $K_1(z)/K_2(z)$ is the thermally averaged dilation factor~\cite{Buchmuller:2004nz}. 
We note that $\Gamma_i$ is a function of $z$ because the mass of $N_i$ depends on temperature. 

$S_i$ is defined as $S_i = \Gamma_S^i/(Hz)$, where $\Gamma_S^i$ is the total scattering rate of $N_i$ given by
\begin{align}
\label{Sterm}
\Gamma_S^i = \frac{T}{64 \pi^4 n_{N_i}^{\text{eq}}} \sum_{N_i f \to f^\prime f^{\prime \prime}} \int_{s_\mathrm{min}^{}}^\infty ds \; \hat{\sigma} \sqrt{s} K_1(\sqrt{s}/T), 
\end{align}
where $n_{N_i}^{\text{eq}}$ is the equilibrium density of $N_i$, and $\sum_{N_i f \to f^\prime f^{\prime \prime}}$ represents the sum of all the scattering processes including $N_i$ in the initial state described in Sec.~\ref{sec: scattering}, both the $\Delta L = +1$ and $-1$ processes. $\hat{\sigma}$ is the reduced cross section for each process defined as~\cite{Plumacher:1996kc} 
\begin{equation}
\hat{\sigma} = \frac{8}{s}\big((p_1 \cdot p_2)^2 - m_1^2m_2^2\big) \sigma,   
\end{equation}
where $p_{1,2}$ and $m_{1,2}$ are the momenta and the masses of the initial particles, respectively, and $\sigma$ is the cross section for the corresponding process.  
Then, the minimum of the integral region is given by $s_\mathrm{min}= (m_1 + m_2)^2$. 
Since we assume that all fermions except $N_i$ are massless in the scatterings, $s_\mathrm{min}$ and $\hat{\sigma}$ are given by 
\begin{equation}
s_\mathrm{min} = M_i(T), \quad \hat{\sigma} = 2 (s-M_i)^2 \sigma/s, 
\end{equation}
common to all the processes. 
After the EWSB, some scattering processes include the $s$-channel contribution from $Z$ and/or $W^\pm$. 
Thus, if $s_\mathrm{min} = M_i(T) < m_{W/Z}$, then the integral region of $\hat{\sigma}$ includes the $W/Z$ pole, and $S_i$ is resonantly enhanced by the on-shell $W/Z$ bosons unless the temperature is much smaller than $m_W$ and $m_Z$.

The equilibrium number density $N_{N_i}^{\mathrm{eq}}$ of the heavy neutrino $N_i$ in a comoving volume is given by
\begin{equation}
N_{N_i}^{\mathrm{eq}} = \frac{3}{8} \left( \frac{M_i(T)}{T} \right)^2 K_2\left( \frac{M_i(T)}{T} \right).
\end{equation}
For $M_i(T) / T > 100$, it can be approximated by the following expression, which includes a Boltzmann suppression factor:
\begin{align}
\label{NeqApproxi}
N_{N_i}^{\mathrm{eq}} \simeq & \; \frac{3}{8} \sqrt{2 \pi \frac{M_i(T)}{T}} \, e^{-M_i(T)/T}
\\
& \; \times \left( \frac{M_i(T)}{T} + \frac{15}{8} + \frac{105}{128 M_i(T)/T} \right). \nonumber
\end{align}
In our scenario, the temperature-dependent part of $M_i(T)$ is proportional to $T^{3/2}$ for $T_\ast > T > T_{N_i}$, as shown in Eq.~\eqref{Mcase}. 
Because $M_i(T)$ decreases faster than $T$, the heavy neutrino $N_i$ can escape Boltzmann suppression. 
As a result, $N_i$ can be produced again via Eq.~\eqref{Neqnres}, thereby realizing the second leptogenesis scenario.

The washout term from the inverse decay contribution is defined as $W_i = \Gamma_{ID}^{(i)} / 2Hz$, where $\Gamma_{ID}^{(i)}$ is the inverse decay rate of the heavy neutrino $N_i$. For a given decay channel with final particles $f$, the inverse decay rate is given by
\begin{equation}
\Gamma_{ID}^{(i)} = \Gamma_D^{(i)} \dfrac{N_{N_i}^\mathrm{eq}}{\prod\limits_{f} N_{f}^\mathrm{eq}} ,
\end{equation}
where $N_f^\mathrm{eq}$ is the comoving equilibrium number density of the final particle $f$, obtained by counting its degrees of freedom. Similarly, one can obtain the washout term from the scattering contribution. After the EWSB, the washout term is the sum of the contributions from each channel.
\begin{equation}
W_i = \sum_{a} \frac{1}{2} D_i^{(a)}  \dfrac{N_{N_i}^\mathrm{eq}}{ \prod\limits_{f} N_{f}^\mathrm{eq}} + \sum_{b} \frac{1}{2} S_i^{(b)}  \dfrac{N_{N_i}^\mathrm{eq}}{\prod\limits_{f} N_{f}^\mathrm{eq}} ,
\end{equation}
where $D_i^{(a)}$ is the decay term and $S_i^{(b)}$ is the scattering term, with indices $a$ and $b$ denoting each channel.

$\varepsilon_{\alpha \beta}^{(i)}$ in Eq.~(\ref{NBLeqnres}) is the $CP$ asymmetry in the decay of $N_i$. 
It is given by the interference terms of the tree level and one-loop level amplitudes. 
At the leading order of $y_{\ell i}^{}$, there are two kinds of contribution to the $CP$ asymmetry. 
One is from the self-energy diagram of $N_i$ ($\varepsilon_{\alpha \beta}^{S(i)}$), and the other is from the one-loop vertex correction ($\varepsilon_{\alpha \beta}^{V(i)}$).
Here, we consider only $\varepsilon_{\alpha \beta}^{S(i)}$ because we focus on the parameter regions where the mass difference $|M_i - M_j|$ is much smaller than their masses, by which $\varepsilon_{\alpha \beta}^{S(i)}$ is resonantly enhanced. 
Such a leptogenesis scenario is called the resonant leptogenesis~\cite{Pilaftsis:1998pd, Pilaftsis:2003gt}. 

Before the EWSB, $\varepsilon_{\alpha \beta}^{S (i)}$ is given by~\cite{Abada:2006fw, DeSimone:2006nrs}
\begin{align}
\label{epsilon_before}
\varepsilon^{S(i)}_{\alpha \beta} =& \frac{\Gamma_i }{2 M_i \{(y^{\dagger} y)_{ii} \}^2} \frac{ (M_j^2 - M_i^2)M_i^2 }{ (M_j^2 - M_i^2)^2 + M_i^4 \Gamma_j^2 / M_j^2}
\nonumber 
\\
& \times
\sum_{j\neq i} \Bigl\{ 
	 i \bigl[ y_{\alpha i} y^\ast_{\beta j} (y^{\dagger} y)_{ji} - y^{\ast}_{\beta i} y_{\alpha j} (y^{\dagger} y)_{ij} \bigr]
 \frac{M_j}{M_i} \nonumber \\
 	& \quad + i  \bigl[ y_{\alpha i} y^\ast_{\beta j} (y^{\dagger} y)_{ij} - y^\ast_{\beta i} y_{\alpha j} (y^{\dagger} y)_{ji} \bigr]
\Bigr\} ,
\end{align}
where $\Gamma_i = (y^\dagger y)_{ii} M_i/(8\pi)$ is the total decay rate of $M_i$.
On the other hand, after the EWSB, the $CP$ asymmetry is given by just replacing $y$ with $R$ in Eq.~(\ref{epsilon_before})\footnote{Strictly speaking, this formula is valid for the process via the gauge interaction. We need to use $y$, not $R$, for the processes via the Higgs coupling such as $N_i \to h \nu$ [see Eq.~(\ref{eq: N_interactions_afterEWSB})]. However, the difference is negligible in the case of small mass differences, which is discussed in the following [see Eq.~(\ref{epsilon_after2})].};
\begin{align}
\label{epsilon_after}
\varepsilon^{S(i)}_{\alpha \beta} =& 
\frac{\Gamma_i }{ 2 M_i \{(R^\dagger R)_{ii}\}^2} \frac{ (M_j^2 - M_i^2)M_i^2 }{ (M_j^2 - M_i^2)^2 + M_i^4 \Gamma_j^2 / M_j^2}
\nonumber 
\\
& \times
\sum_{j\neq i} \Bigl\{ 
	 i \bigl[ R_{\alpha i} R^\ast_{\beta j} (R^{\dagger} R)_{ji} - R^{\ast}_{\beta i} R_{\alpha j} (R^{\dagger}R)_{ij} \bigr]
 \frac{M_j}{M_i} \nonumber \\
 	& \quad + i  \bigl[ R_{\alpha i} R^\ast_{\beta j} (R^{\dagger} R)_{ij} - R^{\ast}_{\beta i} R_{\alpha j} (R^{\dagger} R)_{ji} \bigr]
\Bigr\} .
\end{align}
We can use this formula even when the dominant decay mode is a three-body decay.
By using $R_{\alpha i} = y_{\alpha i} v / (\sqrt{2}M_i)$, we obtain 
\begin{align}
\label{epsilon_after2}
\varepsilon^{S(i)}_{\alpha \beta} = & 
\frac{\Gamma_i}{ 2 M_i \{(yy^\dagger)_{ii}\}^2} \frac{ (M_j^2 - M_i^2)M_i^2 }{ (M_j^2 - M_i^2)^2 + M_i^4 \Gamma_j^2 / M_j^2}
\nonumber 
\\
& \times
\sum_{j\neq i} \left( \frac{ M_i }{ M_j } \right)^2 
\Bigl\{ 
	 i \bigl[ y_{\alpha i} y^\ast_{\beta j} (y^{\dagger} y)_{ji} - y^{\ast}_{\beta i} y_{\alpha j} (y^{\dagger} y)_{ij} \bigr]
 \frac{M_j}{M_i} \nonumber \\
 	& \quad + i  \bigl[ y_{\alpha i} y^\ast_{\beta j} (y^{\dagger} y)_{ij} - y^{\ast}_{\beta i} y_{\alpha j} (y^{\dagger} y)_{ji} \bigr] 
\Bigr\} .
\end{align}
Thus, the difference from the $CP$ asymmetry before the EWSB is only the factor $(M_i/M_j)^2$.
In this paper, we focus on the parameter region where $|M_i - M_j| \simeq \mathcal{O}(\Gamma_j) \ll M_i, M_j$ to obtain large lepton asymmetry with small Yukawa couplings. 
In this case, the correction factor is almost 1, and we can use the same formula for $\varepsilon_{\alpha \beta}^{S (i)}$ at all the eras in the early Universe. 

The similarity of the formula in Eqs.~\eqref{epsilon_before} and \eqref{epsilon_after} is not accidental. 
It is because the absorptive part of the self energy, which is necessary for the $CP$ asymmetry in the decay, is related to the total decay rate by the optical theorem. 
We present the proof in Appendix~\ref{sec: app_CPasym}.  

To discuss the qualitative behavior of the second leptogenesis, consider the following simplified version of the density matrix equation in Eqs.~\eqref{Neqnres}--\eqref{NBLeqnres} with a single generation of the heavy neutrino.
\begin{align}
\frac{\mathrm{d} N_{N_1}}{\mathrm{d} z} &= - (D + S) \bigl(N_{N_1} - N_{N_1}^\mathrm{eq}\bigr), \label{Boltz1} \\
\frac{\mathrm{d} N_{B-L}}{\mathrm{d} z} &= - \varepsilon D \bigl(N_{N_1} - N_{N_1}^\mathrm{eq}\bigr) - W N_{B-L}. \label{Boltz2}
\end{align}
By substituting the first equation into the second, the evolution of the $B-L$ asymmetry is given by~\cite{Buchmuller:2004nz}
\begin{align}
\label{NBmLevol}
N_{B-L}(z) = & N_{B-L}^i e^{-\int_{z_i}^z dz' W(z')} 
\\ & - \int_{z_i}^{z} dz' \frac{\varepsilon D}{D+S} \frac{\mathrm{d} N_{N_1}}{\mathrm{d} z'} e^{-\int_{z'}^z dz'' W(z'')} , \nonumber
\end{align}
where $N^i_{B-L}$ is the initial value of the $B-L$ asymmetry at $z_i$ for each period.
The $CP$ asymmetry $\varepsilon$ also appears within the integration, since it varies over time due to the mass variation in our scenario.
Here, we divide the analysis into three periods based on their distinct behaviors: the first leptogenesis before EWSB, the secondary production of $N_{N_1}$ after EWSB and mass variation, and the out-of-equilibrium decay of $N_{1}$ in the late Universe.

The behavior of leptogenesis and the generated lepton asymmetry differs depending on whether the washout is weak or strong. This is determined by the ratio between the production rate $\Gamma_{\mathrm{prod}} \propto y^2 M(T) \propto M_{0} M(T)$ and the Hubble rate $H \propto T^2$. Here we have suppressed the index for the heavy neutrino generation.
For the first leptogenesis with $M(T) \simeq M_{\ast}$ and $T \simeq M_{\ast}$, the ratio is given by
$\Gamma_{\mathrm{prod}}/H \; ( = zD + zS ) \propto M(T) M_{0} /T^2 \simeq M_{0}/M_{\ast} \ll 1$, 
which corresponds to the weak washout regime. Because the production rate is suppressed, the number density $N_{N_1}$ of the heavy neutrino cannot follow the equilibrium density $N_{N_1}^{\mathrm{eq}}$.
During the early epoch, the heavy neutrino mass is fixed at $M_{\ast}$, so the evolution of the lepton asymmetry behaves similarly to conventional leptogenesis under the weak washout regime.

After the start of the mass variation, the equilibrium density of $N_1$ becomes large again because $M(T)$ is not significantly greater than $T$. This leads to $N_1$ being produced once more.  
The mass variation effect can be inserted by replacing $z \rightarrow z \, M(T)/M_0$.  
Moreover, after EWSB, the scattering term $S$ becomes momentarily dominant due to changes in the decay channels and the resonant enhancement of scattering.  
For the second leptogenesis in the late Universe, the ratio between the production rate and the Hubble rate is given by $\Gamma_{\mathrm{prod}}/H \propto M(T)\,M_{0}/T^2\big|_{T=M_{0}} \simeq 1$, which corresponds to the strong washout regime and leads to a large lepton asymmetry.

In the final stage, the heavy neutrino disappears via out-of-equilibrium decay.
Because the decay rate is suppressed by the Fermi constant in Eqs.~\eqref{3nu}--\eqref{lqq}, this decoupling occurs later.
The temperature-dependent part becomes negligibly small compared with the bare mass term, so the heavy neutrino mass is fixed in the late Universe.
The scattering term $S$ and washout term $W$ are small in this epoch, and as a result, Eq.~\eqref{NBmLevol} takes the simple form $N_{B-L}(z) \simeq N_{B-L}^i - \varepsilon N_{N_1}(z)$.

The realization of the second leptogenesis scenario with a specific parameter setup will be presented in the next section by numerically solving the density matrix equation under resonant leptogenesis with all three generations of $N_i$.

\section{Numerical demonstration of the second leptogenesis}
\label{Estimation}

In order to achieve large $CP$ asymmetry in the decay, we focus on a scenario where the masses of the three heavy neutrinos are nearly degenerate. This configuration resonantly enhances the $CP$ asymmetry, and it can reach the maximal value independent of the size of the Yukawa coupling~\cite{Pilaftsis:1998pd, Pilaftsis:2003gt}.
It enables us to obtain a large $B-L$ asymmetry with relatively small masses of $N_i$ in the current Universe.

At zero temperature, $\Delta M^0_{12} \equiv M_{02} - M_{01}$ and $\Delta M^0_{13} \equiv M_{03} - M_{01}$ for the bare mass term are set to meet the resonance condition, $|M_{0i} - M_{0j}| = \Gamma_{0j}/2$, where $\Gamma_{0j}$ is the decay rate of $N_j$ at zero temperature. This configuration maximizes the $CP$ asymmetry, leading to a large LAU after the second leptogenesis.
In contrast, the mass differences at higher temperatures than $T_\ast$, $\Delta M^\ast_{12} \equiv M_{\ast 2} - M_{\ast 1}$ and $\Delta M^\ast_{13} \equiv M_{\ast 3} - M_{\ast 1}$ are determined to reproduce the observed BAU by the first leptogenesis.
They are taken to be small but do not satisfy the resonant condition, i.e., the obtained $CP$ asymmetry is small compared to that in the current Universe.
We note that the temperature-dependent mass enables the significant difference in the size of the $CP$ asymmetry at high temperatures and at zero temperature. This plays a crucial role in explaining the large discrepancy between the BAU and LAU.

The $CP$-violating sources arise from the imaginary parts of the Yukawa matrix. The Casas-Ibarra parametrization \cite{Casas:2001sr} of the Yukawa matrix is expressed as 
\begin{align}
y_{\ell i} = \frac{\sqrt{2}i}{v} U_{\ell a} m_a^{1/2} O_{ai} M_i^{1/2} ,
\end{align}
where $\ell = e, \mu, \tau$, $a, i = 1,2,3$, and $U_{\ell a}$ is the PMNS matrix element.
We follow the notation in Ref.~\cite{ParticleDataGroup:2022pth} for Majorana $CP$-violating phases $\alpha_1$ and $\alpha_2$ in the PMNS matrix. The complex orthogonal matrix $O$, including three complex phases $\omega_i$ ($i=1,2,3$), is defined as \cite{Moffat:2018wke}
\begin{align}
O = &  
\begin{pmatrix}
1 & 0 & 0
\\
0 & \cos \omega_1 & \sin \omega_1 
\\
0 & -\sin \omega_1 & \cos \omega_1 
\end{pmatrix}
\begin{pmatrix}
\cos \omega_2 & 0 & -\sin \omega_2
\\
0 & 1 & 0 
\\
\sin \omega_2 & 0  & \cos \omega_2 
\end{pmatrix} \nonumber
\\
& \times
\begin{pmatrix}
\cos \omega_3 & \sin \omega_3 & 0
\\
- \sin \omega_3 & \cos \omega_3 & 0 
\\
0 & 0  & 1 
\end{pmatrix}.
\end{align}

\begin{figure}[tb]
\centering
\includegraphics[width=0.49\textwidth]{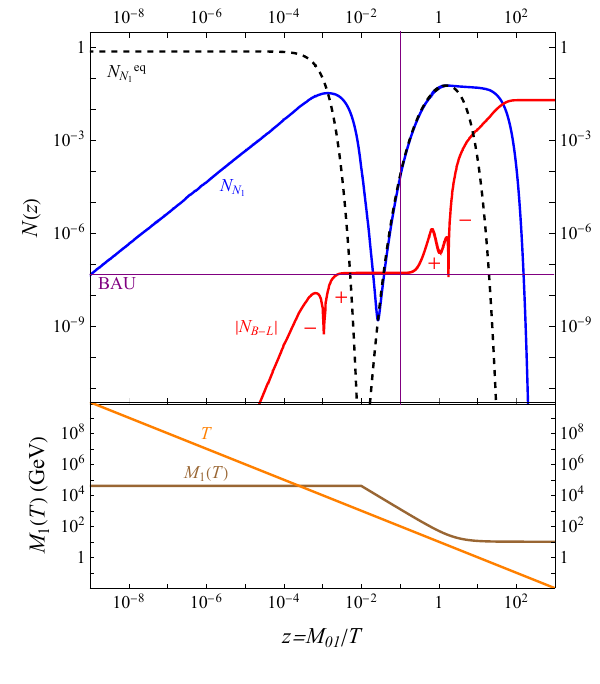}
\caption{
The heavy neutrino number density $N_{N_1}$ and the $B-L$ asymmetry $\left|N_{B-L}\right|$ are plotted against $z = M_{01}/T$. Here, the input parameters are set as $M_{01} = 10$ GeV, $M_{\ast 1} = 40$ TeV, $T_\ast = 1$ TeV ($m_\phi \simeq 1.4 \times 10^{-3}$ eV), with other values provided in the text. The horizontal purple line represents the value of $|N_{B-L}|$ required to explain all the amount of the observed BAU. The vertical purple line shows the $T_{\text{EW}} = 100$ GeV, around which the sphaleron decoupling is expected to occur. The decoupling happens after the first leptogenesis ($z \simeq 10^{-3}$) but before the second one ($z \simeq 1$). The sign of $N_{B-L}$ is indicated near the curve.
The heavy neutrino mass $M_1(T)$ and the temperature $T$ are plotted together.
}
\label{FinalPlot}
\end{figure}

Figure~\ref{FinalPlot} illustrates the behavior of $N_{N_1}$ and $|N_{B-L}| = |\sum_\alpha N_{\alpha \alpha}|$ during the first and second phases of leptogenesis.
For this demonstration, the input parameters are set to $M_{01} = 10$ GeV, $M_{\ast 1} = 40$ TeV, and $T_\ast = 1$ TeV ($m_\phi \simeq 1.4 \times 10^{-3}$ eV).
This input avoids the theoretical and experimental constraints discussed in Sec.~\ref{Model}.
The free parameters in the Yukawa matrix are chosen as follows. $m_{\nu_1} = 10^{-3}$ eV, assuming the normal hierarchy of the active neutrino masses. The angles are set to $\alpha_1 = \alpha_2 = 0$ and $\omega_1 = \omega_2 = 0$, with $\omega_3 = 0.2e^{i\pi/4}$ as the only nonzero angle.
The mass differences in the current Universe $\Delta M^0_{12}$ and $\Delta M^0_{13}$ satisfy the resonance condition, $\Delta M^0_{12} = \Gamma_{02}/2>0$ and $\Delta M^0_{13} = \Gamma_{03}/2>0$.
The mass difference at high temperatures are set to be $\Delta M^\ast_{12} = -2.0 \times 10^{-9}$ eV and $\Delta M^\ast_{13} = -2.0 \times 10^{-8}$ eV. 
The mass hierarchy among $N_i$ is different at high and low temperatures to obtain the desired sign of the $CP$ asymmetry, resulting in the $B-L$ generation with the correct signs for each corresponding cosmological period.

For simplicity, we assume that the sphaleron decoupling occurs instantaneously at $T=T_\mathrm{EW} = 100~\mathrm{GeV}$, allowing us to directly use the value of $N_{B-L}$ at $T_\mathrm{EW}$ to evaluate the baryon asymmetry. On the other hand, the final lepton asymmetry is determined by the value at $T \ll M_{01}$.

The baryon-to-photon ratio and lepton-to-photon ratio are evaluated by
\begin{align}
    \eta_B = \left. \frac{a_{\text{sph}} N_{B-L}}{f}   \right|_{T=T_\mathrm{EW}} , \quad
    \eta_L = \left. \frac{N_{B-L}}{f'} \right|_{T = T_\mathrm{BBN}} , 
\end{align}
where $T_\mathrm{BBN} \simeq 0.1~\mathrm{MeV}$ is the typical temperature of the BBN, $a_\mathrm{sph} = 28/79$~\cite{Kuzmin:1985mm} is a sphaleron conversion factor, and $f = 1232/43$ and $f^\prime = 1012/43$ are the photon dilution factor~\cite{Buchmuller:2002rq} at the temperature of the first and second lepgogenesis, respectively.
In the evaluation of $f$ and $f^\prime$, we include all three heavy neutrino species because their decoupling behavior is almost the same in our benchmark point~\cite{ChoeJo:2023cnx}.
The benchmark point can reproduce the observed BAU within $1~\sigma$, $\eta_B \simeq 6.14\times 10^{-10}$, and simultaneously achieve $\eta_L^{} \simeq 8.5 \times 10^{-4}$. Although this is significantly larger than $\eta_B$, it is about one order of magnitude lower than the value from EMPRESS data.

\begin{figure}[tb]
\centering
\includegraphics[width=0.49\textwidth]{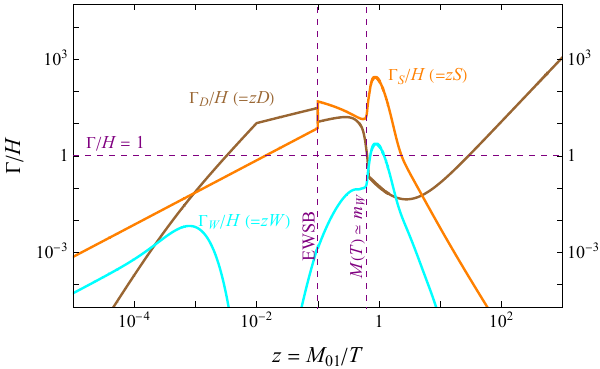}
\caption{
The decay, scattering, and washout terms over the expansion factor $z$, using the same parameter values as in Fig.~\ref{FinalPlot}. The dominant decay channels change at the point of EWSB and when $M(T) \simeq m_{Z}$ (indicated by dashed purple lines), causing kinks to appear in $\Gamma_D/H (= zD)$. The ratio for the scattering rate, $\Gamma_S/H (= zS)$, exhibits resonance near $M(T) \simeq m_W$. Only the rates for the first heavy neutrino, $N_1$, are shown here. 
}
\label{RatePlot}
\end{figure}

Figure~\ref{RatePlot} compares the decay, scattering, and washout rates of the first heavy neutrino $N_1$ over the expansion rate $H$ with respect to $z = M_{01}/T$. 
Large discontinuities in $\Gamma_D/H (= zD)$ and $\Gamma_S/H (= zS)$ curves appear at $T = T_\mathrm{EW}$ because the weak interaction of $N_i$ is switched on. 
A lot of new scattering channels open, while the decay rate is suppressed by the mass of the gauge and Higgs bosons as discussed in Sec.~\ref{DecayScattering}. 
The discontinuities are relaxed if we consider the temperature-dependent VEV profiles. 

The washout effect does not seem to have such a discontinuity at $T = T_\mathrm{EW}$. However, in each contribution from the inverse decays and the scatterings, there are discontinuities due to those in $\Gamma_D$ and $\Gamma_S$. Since the change of $\Gamma_D$ and $\Gamma_S$ have opposite signs, the discontinuity in the sum of them is relaxed, and it looks smooth. This cancellation of discontinuity would depend on the choice of the model parameters.

There is another discontinuity in $\Gamma_D$ at $T \simeq m_W^{}$, which is caused by the change of the dominant decay modes due to the kinematics.
As explained in Sec.~\ref{DecayScattering}, we use the Fermi interaction for the three-body decay at low temperatures and do not fully take into account the effect of the off-shell gauge bosons. Thus, there is a gradual discontinuity at the threshold of the decay into the gauge bosons.
On the other hand, $\Gamma_S$ does not have such a discontinuity because we do not use the Fermi interaction.

For a while after the EWSB, the scattering processes are in equilibrium ($\Gamma_S > H$), and the heavy neutrinos are thermalized. 
The scattering rate increases furthermore below the temperature such that $M(T) \simeq m_W^{}$ or $m_Z$ due to the resonance enhancement in the cross section via the on-shell gauge bosons as discussed in Sec.~\ref{DecayScattering}. 
The decoupling of the scatterings occurs at $z \simeq 2$. 
From the decoupling of the scattering to a period at $z \simeq 30$, both the scatterings and the decays are ineffective, and the number densities of $N_i$ are frozen. 
This makes the contours of $N_i$ (the blue line) flat for $2 \lesssim z \lesssim 30$ in Fig.~\ref{FinalPlot}.
After $z \simeq 30$, the decay rate exceeds the cosmic expansion rate, and the large lepton asymmetry is generated by the second leptogenesis. $N_i$ are mostly gone at $z \simeq 10^2$, i.e., $T \simeq 100~\mathrm{MeV}$. Thus, we expect that it avoids constraints from BBN and CMB.

In our benchmark point, $N_i$ are thermalized before the second leptogenesis, and the washout effect is ineffective at that time. 
Thus, the final lepton asymmetry is almost entirely determined by only their equilibrium number density and the size of the $CP$ asymmetry in the decay like the strong washout case in vanilla leptogenesis~\cite{Davidson:2008bu}. 
On the other hand, $N_i$ are not thermalized before the first leptogenesis. The amount of the baryon asymmetry depends on various conditions before $T = T_\mathrm{EW}$, for example, the initial condition about the population of $N_i$, like the weak washout case~\cite{Davidson:2008bu}.

\begin{figure}[tb]
\centering
\includegraphics[width=0.49\textwidth]{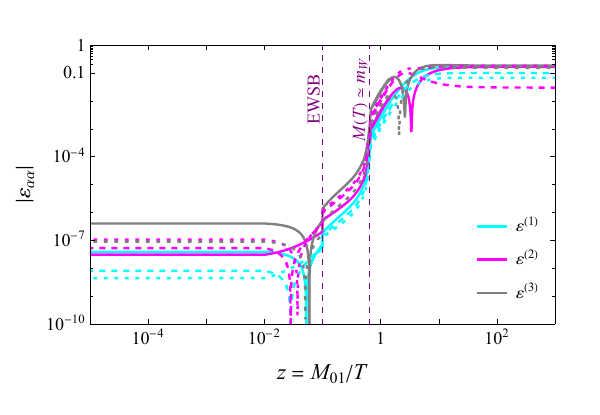}
\caption{
The temperature dependence of the absolute values of each $\varepsilon^{(i)}_{\alpha\alpha}$ is shown using the same parameter values as in Fig.~\ref{FinalPlot}. 
Light blue, pink, and gray curves indicate $i=1,2,3$, respectively.
Solid, dashed, and small dashed curves correspond to the $e$, $\mu$, and $\tau$ components, respectively.
When the dominant decay channel is altered at vertical lines, some components show discontinuities.
$\varepsilon^{(i)}_{\alpha\alpha}$ are small in the early Universe but take almost the maximal value in the late Universe by satisfying the resonance condition. This aids the generation of a large lepton asymmetry in the second leptogenesis epoch.}
\label{EpsPlot}
\end{figure}

The behavior of the diagonal terms of the $CP$ asymmetry in the decay, $\varepsilon^{(i)}_{\alpha\alpha}$, is depicted in Fig.~\ref{EpsPlot} with the same input values.
We note that its magnitude $|\varepsilon_{\alpha\alpha}^{(i)}|$ is small in the early Universe where the resonant condition is not satisfied, and it gets amplified as the Universe expands and the mass differences match the resonant condition for leptogenesis. As previously mentioned, this can provide large asymmetry generation during the second leptogenesis epoch, thereby enlarging the discrepancy between baryon and lepton asymmetry.

The sign of the $N_{B-L}$ changes during the cosmic expansion as represented in Fig.~\ref{FinalPlot}, depending on the sign and magnitude of the coefficients in Eq.~(\ref{NBLeqnres}). The first sign flip near $z\simeq10^{-3}$ is due to the change of the sign of $N_{N_i}-N_{N_i}^{\text{eq}}$, while the second flip near $z\simeq2$ is generated by the sign flip of the asymmetry factor $\varepsilon^{(i)}_{\alpha\alpha}$ itself as shown in Fig.~\ref{EpsPlot}. 

We note that there is a distinctive bump in the $N_{B-L}$ contour in Fig.~\ref{FinalPlot} just before the second sign flip in $B-L$ asymmetry. It is caused by the balance of the effects of the decay and washout processes as described below. As shown in Fig.~\ref{RatePlot}, after the EWSB, the decay is more effective than the washout processes until the temperature at which $M(T) \simeq m_W$. $B-L$ asymmetry is generated via the decay, and $|N_{B-L}|$ increases. 
Around the point $z = 1$, the rate of the washout effect exceeds one due to the effect of the on-shell gauge bosons in the scatterings. 
On the other hand, the decay rate is suppressed because only the three-body decays are kinematically allowed. 
Thus, $B-L$ asymmetry is partly washed out in this region, and $|N_{B-L}|$ decreases. 
Soon after $z \simeq 1$, the washout effect starts to decrease, and it becomes smaller than the decay rate at $z \simeq 1.3$. Therefore, $|N_{B-L}|$ starts to increase again at this point. 

The behavior of leptogenesis is influenced by changes in several parameters: the bare mass term $M_0$, the mass-varying term $M_\ast$, and the temperature $T_\ast$ at which mass variation begins. When $M_0$ gets smaller, the variable $z$ becomes small at the same temperature, shifting the curves in Fig.~\ref{FinalPlot} to the left. As a result, the heavy neutrino $N$ decays more slowly at later times with such a small $M_0$. The mass-varying term $M_\ast$ affects both the production and decay of $N$ in the early epoch: lower values of $M_\ast$ increase production, while higher values decrease it.
If $T_{\ast}$ gets lower, mass variation occurs later, meaning that the mass $M(T)$ remains large at $T \simeq M_0$, causing the decrease of the production of $N$.

The number density $N_{N_1}$ of the lightest heavy neutrino is $\mathcal{O}(0.1)$ near $z \simeq 1$. The lepton asymmetry might be amplified if this number density is larger. This increase could be possible if $T_{\ast}$ is high, allowing the mass to start decreasing earlier. However, maintaining the same baryon asymmetry under these conditions would require a higher electroweak temperature $T_{\text{EW}}$. One method to increase $T_{\text{EW}}$ involves introducing a new scalar field $S$ \cite{Baldes:2018nel}, where the VEV of $S$, denoted as $\langle S \rangle$, sets a new EWSB scale significantly higher than $v$. This adjustment leads to earlier sphaleron decoupling compared to the SM.

Another strategy to enhance the lepton asymmetry involves reducing the mass scale of $N$, which would delay the end of the second leptogenesis phase. However, such a low mass for the heavy neutrinos might face strong constraints from BBN or CMB observations since they are thermalized before the leptogenesis.

\section{Summary and outlook}
\label{Summary}

The Majorana mass scale is a crucial parameter in leptogenesis. If the mass of the heavy neutrino changes over cosmic history, it can significantly modify the behavior of leptogenesis. Particularly, if the mass decreases more rapidly than the temperature, the heavy neutrino may be produced again, allowing leptogenesis to occur once more upon their subsequent decay. Such a scenario can be realized through coupling with wave dark matter.

The benefit of introducing the second leptogenesis comes from the existence of two distinct phases of matter-antimatter asymmetry generation. If the sphaleron decoupling occurs between the first and second phases of leptogenesis, the asymmetry generated in the second phase cannot be converted into a baryon number. In this scenario, one can expect that the baryon asymmetry is fixed after the first phase of leptogenesis, while the lepton asymmetry can be further enhanced during the second phase. This mechanism can potentially explain the large discrepancy between the baryon and lepton asymmetries. 

This paper presents a detailed analysis of the weak interaction processes of the heavy neutrino across a wide mass range. Under conditions of temperature-dependent mass, the relevant decay channels vary across different cosmic periods. After the EWSB, the heavy neutrinos decay into massive on-shell bosons and leptons. However, these decay channels cease when the mass of the heavy neutrino falls below that of these bosons. This alteration in decay channels introduces modification in the behavior of leptogenesis. Additionally, scattering via the gauge boson in the $s$ channel can achieve resonance as the mass of the heavy neutrino passes through the $W/Z$ boson mass. Our model also allows the $CP$ asymmetry to vary over time, which can lead to increased asymmetry generation in the later stages.

In our scenario, both the bare masses and the dark matter coupling strength of all three heavy neutrinos are nearly degenerate. This calls for a model with a new mechanism for a softly broken flavor symmetry, which could be the subject of future research.

The final lepton asymmetry estimated at our benchmark point is remarkably sizable, yet it is somewhat smaller than the reported EMPRESS value. While this could potentially be resolved with an optimized benchmark point, it could also be addressed by introducing elements of new physics, such as a higher temperature for EWSB. Such an adjustment could increase the equilibrium number density of the heavy neutrino and, consequently, enhance the lepton asymmetry further.

\begin{acknowledgments}
This work was supported in part by the National Research Foundation of Korea (Grant No. RS-2024-00352537).
\end{acknowledgments}

\appendix

\section{CROSS SECTIONS FOR ELECTROWEAK SCATTERING OF STERILE NEUTRINOS}
\label{Appendix}

Some common factors for the scattering cross sections are defined for convenience:
\begin{align}
& \sigma_0 \equiv \frac{\abs{y_{\ell i}}^2 v^2}{m_N^2(s-m_N^2) } ,
\\
& f_s(m_s) \equiv \frac{(s-m_N^2)(2s+m_N^2)}{(s-m_s^2)^2+m_s^2\Gamma_s^2} , \\
& f_{t1}(m_t) \equiv \frac{s(s-m_N^2)}{m_t^2(s+m_t^2-m_N^2)} , \\
& f_{t2}(m_t) \equiv 2+\frac{s}{m_t^2}-\left(1+\frac{s+2m_t^2}{s-m_N^2}\right)\log\left(1+\frac{s-m_N^2}{m_t^2}\right),
\end{align}
\begin{align}
& f_{st}(m_s,m_t) \equiv \frac{s-m_s^2}{(s-m_s^2)^2+m_s^2\Gamma_s^2}  
\\
& \hspace{50pt} \times \left(3s+2m_t-m_N^2 \right.  \nonumber
\\
& \hspace{50pt} \; \left. -2(s+m_t^2)\Big(1+\frac{m_t^2}{s-m_N^2}\Big)\log\Big(1+\frac{s-m_N^2}{m_t^2}\Big)\right) , \nonumber \\
& f_{tu}(m_t,m_u) \equiv \frac{s}{s+m_t^2+m_u^2-m_N^2} \nonumber \\
& \hspace{50pt} \times \left(\log\Big(1+\frac{s-m_N^2}{m_t^2}\Big) +\log\Big(1+\frac{s-m_N^2}{m_u^2}\Big)\right) .
\end{align}
$f_s$ appears in $s$-channel diagrams, $f_{t1}$ and $f_{t2}$ appear in $t$ or $u$-channel diagrams. $f_{st}$ corresponds to the cross term of the $s$ and $t(u)$ channels, and $f_{tu}$ corresponds to the cross term of the $t$ and $u$ channels. The $Z$ or $W$ boson mass is inserted into $m_s$ and $m_t$ depending on each channel. We will denote $\sin^2\theta_W \equiv s_W^2$ and refer to up- and down-type quarks as $u, d$ collectively. The formulas for the scattering cross sections for a given $\ell$ are the following.

For the channels with a single $s$-channel diagram:

\begin{enumerate}[(a)]

\item $N_i\nu_\ell \rightarrow f\bar{f}$

\begin{align}
\sum_u\sigma(N_i\nu_\ell \rightarrow u\bar{u}) &= \frac{g_Z^4}{3072\pi} \sigma_0 f_s(m_Z) 
\nonumber \\
& \quad 
\times 6 \left(1-\frac{8}{3}s_W^2+\frac{32}{9}s_W^4 \right) ,
\\
\sum_d\sigma(N_i\nu_\ell \rightarrow d\bar{d}) &= \frac{g_Z^4}{3072\pi} \sigma_0 f_s(m_Z) \nonumber \\
& \quad 
\times 9 \left(1-\frac{4}{3}s_W^2+\frac{8}{9}s_W^4\right) ,
\\
\sum_{\ell\neq\ell'}\sigma(N_i\nu_\ell \rightarrow \ell'\bar{\ell}') &= \frac{g_Z^4}{3072\pi} \sigma_0 f_s(m_Z) \nonumber \\
& \quad 
\times 2 \left(1-4s_W^2+8s_W^4 \right) ,
\\
\sum_{\ell\neq\ell'}\sigma(N_i\nu_\ell \rightarrow \nu_{\ell'}\bar{\nu}_{\ell'}) &= \frac{2 g_Z^4}{3072\pi} \sigma_0 f_s(m_Z) .
\end{align}

\item $N_i\ell \rightarrow d\bar{u}/\ell'\bar{\nu}_{\ell'}$
\begin{align}
\sum_{u,d}\sigma(N_i\ell \rightarrow d\bar{u}) &= \frac{g_W^4}{256\pi} \sigma_0  f_s(m_W) \cdot \sum_{u,d}\abs{V_{ud}}^2 ,
\\
\sum_{\ell\neq\ell'}\sigma(N_i\ell \rightarrow \ell'\bar{\nu}_{\ell'}) &= \frac{g_W^4}{384\pi} \sigma_0 f_s(m_W) .
\end{align}

For the channels with a single $t$- or $u$-channel diagram with only external leptons:

\item $N_i\bar{\nu}_{\ell'} \rightarrow \bar{\nu}_\ell \bar{\nu}_{\ell'} ~(\ell \neq \ell')$
\begin{align}
\sum_{\ell\neq\ell'}\sigma(N_i\bar{\nu}_{\ell'} \rightarrow \bar{\nu}_\ell \bar{\nu}_{\ell'}) &= \frac{g_Z^4}{256\pi} \sigma_0 f_{t1}(m_Z) .
\end{align}

\item $N_i\nu_{\ell'} \rightarrow \ell'\bar{\ell} ~(\ell \neq \ell')$
\begin{align}
\sum_{\ell\neq\ell'}\sigma(N_i\nu_{\ell'} \rightarrow \ell'\bar{\ell}) &= \frac{g_W^4}{64\pi} \sigma_0 f_{t2}(m_W) .
\end{align}

\item $N_i\nu_{\ell'} \rightarrow \nu_{\ell'}\bar{\nu}_\ell ~(\ell \neq \ell')$
\begin{align}
\sum_{\ell\neq\ell'}\sigma(N_i\nu_{\ell'} \rightarrow \nu_{\ell'}\bar{\nu}_\ell) &= \frac{g_Z^4}{256\pi} \sigma_0  f_{t2}(m_Z) .
\end{align}

\item $N_i\ell' \rightarrow \ell'\bar{\nu}_\ell ~(\ell \neq \ell')$
\begin{align}
\sum_{\ell\neq\ell'}\sigma
(N_i\ell' \rightarrow \ell'\bar{\nu}_\ell) &= \frac{g_Z^4}{256\pi} \sigma_0  \Big(4s_W^4 f_{t1}(m_Z) \nonumber \\
& \quad 
+ (1-4s_W^2+4s_W^4) f_{t2}(m_Z)\Big) .
\end{align}

\item $N_i\bar{\ell}' \rightarrow \bar{\ell}'\bar{\nu}_\ell ~(\ell \neq \ell')$
\begin{align}
\sum_{\ell\neq\ell'}\sigma(N_i\bar{\ell}' \rightarrow \bar{\ell}'\bar{\nu}_\ell) &= \frac{g_Z^4}{256\pi} \sigma_0  
\Big((1-4s_W^2+4s_W^4) f_{t1}(m_Z) \nonumber \\
& \quad 
+ 4s_W^4 f_{t2}(m_Z)\Big) .
\end{align}

\item $N_i\bar{\ell}' \rightarrow \bar{\ell}\nu_{\ell'} ~(\ell \neq \ell')$
\begin{align}
\sum_{\ell\neq\ell'}\sigma(N_i\bar{\ell}' \rightarrow \bar{\ell}\nu_{\ell'}) &= \frac{g_W^4}{64\pi} \sigma_0 f_{t1}(m_W) .
\end{align}

For the channels with a  single $t$- or $u$-channel diagram including external quarks:

\item $N_i q \rightarrow \bar{\nu}_\ell q$
\begin{align}
\sum_u\sigma(N_i u \rightarrow \bar{\nu}_\ell u) &= \frac{g_Z^4}{512\pi} \sigma_0 \cdot 2  \left(\frac{16}{9} s_W^4 f_{t1}(m_Z) \right. \nonumber \\
& \quad \left.
+ (1-\frac{8}{3}s_W^2+\frac{16}{9}s_W^4) f_{t2}(m_Z)\right)  ,
\\
\sum_d\sigma(N_i d \rightarrow \bar{\nu}_\ell d) &= \frac{g_Z^4}{512\pi} \sigma_0 \cdot 3  \left(\frac{4}{9} s_W^4 f_{t1}(m_Z) 
\right. \nonumber \\
& \quad \left. 
+ (1-\frac{4}{3}s_W^2+\frac{4}{9}s_W^4) f_{t2}(m_Z)\right) .
\end{align}

\item $N_i \bar{q} \rightarrow \bar{\nu}_\ell \bar{q}$
\begin{align}
\sum_u\sigma(N_i \bar{u} \rightarrow \bar{\nu}_\ell \bar{u}) &= \frac{ g_Z^4}{512\pi} \sigma_0 
\nonumber \\
& \quad
\times 2 \left((1-\frac{8}{3}s_W^2+\frac{16}{9}s_W^4) f_{t1}(m_Z) 
\right. \nonumber \\
& \quad \left. 
+ \frac{16}{9} s_W^4 f_{t2}(m_Z)\right) , \\
\sum_d\sigma(N_i \bar{d} \rightarrow \bar{\nu}_\ell \bar{d}) &= \frac{g_Z^4 }{512\pi} \sigma_0 
\nonumber \\
& \quad
\times 
3 \left((1-\frac{4}{3}s_W^2+\frac{4}{9}s_W^4) f_{t1}(m_Z) 
\right. \nonumber \\
& \quad \left. 
+ \frac{4}{9} s_W^4 f_{t2}(m_Z)\right) .
\end{align}

\item $N_i u \rightarrow \bar{\ell} d$
\begin{align}
\sum_{u,d,\ell}\sigma(N_i u \rightarrow \bar{\ell} d) &= \frac{g_W^4}{128\pi} \sigma_0 f_{t2}(m_W) \cdot \sum_{u,d}\abs{V_{ud}}^2 ,
\end{align}

\item $N_i \bar{d} \rightarrow \bar{\ell} \bar{u}$
\begin{align}
\sum_{u,d,\ell}\sigma(N_i \bar{d} \rightarrow \bar{\ell} \bar{u}) &= \frac{g_W^4}{128\pi} \sigma_0  f_{t1}(m_W) \cdot \sum_{u,d}\abs{V_{ud}}^2 .
\end{align}

For the diagrams including both $s$ and $t(u)$ channels:

\item $N_i \nu_\ell \rightarrow \ell\bar{\ell}$
\begin{align}
\sigma(N_i \nu_\ell \rightarrow \ell\bar{\ell}) = & \frac{1}{32\pi} \sigma_0   
\left(\frac{g_Z^4}{96} (1-4s_W^2+8s_W^4) f_s(m_Z)  \right. \nonumber
\\
& \left. + \frac{g_W^4}{4} f_{t2}(m_W) - \frac{g_Z^2g_W^2}{8} f_{st}(m_Z,m_W)\right) .
\end{align}

\item $N_i \nu_\ell \rightarrow \nu_\ell \bar{\nu}_\ell$
\begin{align}
\sigma(N_i \nu_\ell \rightarrow \nu_\ell \bar{\nu}_\ell) &= \frac{g_Z^4}{512\pi} \sigma_0  \left(\frac{1}{6}f_s(m_Z) + f_{t2}(m_W) 
\right. \nonumber
\\
& \quad 
+ f_{st}(m_Z,m_W) \bigg). 
\end{align}

\item $N_i \ell \rightarrow \ell \bar{\nu}_\ell$
\begin{align}
\sigma(N_i \ell \rightarrow \ell \bar{\nu}_\ell) = \frac{1}{32\pi} \sigma_0  & \left(  \frac{g_W^4}{24} f_s(m_W) + \frac{g_Z^4}{4} f_{t1}(m_Z) \right. \nonumber \\
& \quad \left. + \frac{g_Z^4}{16} (1-4s_W^2+4s_W^4) f_{t2} 
\right. \nonumber
\\
& \quad \left.
- \frac{g_Z^2g_W^2}{8} f_{st}(m_W,m_Z)\right) .
\end{align}

For the diagrams including both $t$ and $u$ channels:

\item $N_i \bar{\nu}_\ell \rightarrow \bar{\nu}_\ell\bar{\nu}_\ell$
\begin{align}
\sigma(N_i \bar{\nu}_\ell \rightarrow \bar{\nu}_\ell\bar{\nu}_\ell) &= \frac{ g_Z^4}{512\pi} \sigma_0   \big(f_{t1}(m_Z) + f_{tu}(m_Z,m_Z)\big) .
\end{align}

\item $N_i\bar{\ell} \rightarrow \bar{\ell}\bar{\nu}_\ell$
\begin{align}
\sigma(N_i\bar{\ell} \rightarrow \bar{\ell}\bar{\nu}_\ell) = \frac{1}{32\pi} \sigma_0  & \left( \frac{g_Z^4}{16} (1-4s_W^2+4s_W^4) f_{t1}(m_Z) 
\right. \nonumber
\\
& \quad \left.
+ \frac{g_W^4}{4} f_{t1}(m_W) + \frac{g_Z^4}{4} s_W^4 f_{t2}(m_Z) 
\right. \nonumber
\\
& \quad \left.
- \frac{g_Z^2g_W^2}{4} (1-2s_W^2) f_{tu}(m_W,m_Z)\right) .
\end{align}

\end{enumerate}

\section{THE DERIVATION OF THE GENERAL FORMULA FOR $CP$ ASYMMETRY}
\label{sec: app_CPasym}

In this appendix, we derive the $CP$ asymmetry of the decay of $N_i$ after the EWSB in Eq.~\eqref{epsilon_after}. We consider the decays via the gauge bosons, i.e., $N_i \to \psi_\ell V^{(\ast)}$ and $N_i \to \bar{\psi}_\ell V^{\dagger(\ast)}$, where $\psi_\ell$ is a lepton, and $(\ast)$ indicates that the vector boson $V$ may be offshell. The extension to the case of $N_i \to \nu_{\ell}(\bar{\nu}_{\ell}) h$ is straightforward. We find the difference in their decay rates at the leading order of $y_{\ell i}$. This corresponds to the diagonal terms of the $CP$ asymmetry matrix $\varepsilon^{(i)}$. The nondiagonal terms can be obtained by appropriately replacing the lepton flavor indices.

\subsection{Identities for the 1PI self-energy of Majorana fermions}

Before considering the decay processes, we derive the fundamental relation for the 1PI self-energy of Majorana fermions. We consider the flavored two-point functions of the heavy neutrinos defined as
\begin{equation}
\label{eq: 2point_function}
i S_{ij}(x,y) \equiv \left< T N_i (x) \overline{N}_j (y) \right>,
\end{equation}
where $T$ represents the time-ordered product. In the following, we employ the method for deriving the Feynman rules shown in Ref.~\cite{Denner:1992vza} and use only the two-point functions in the form of Eq. \eqref{eq: 2point_function} even for Majorana fermions.

In the momentum space, the K\"{a}llen-L\'{e}hman representation of $S_{ij}$ is given by~\cite{Grimus:2016hmw}
\begin{align}
\label{eq: KL_Sij}
S_{ij}(\cancel{p}) = & \int_0^\infty \mathrm{d}\mu^2 \Bigl[ \cancel{p} \bigl\{c^L_{ij}(\mu^2) P_L + c^R_{ij}(\mu^2) P_R\bigr\}  
\nonumber
\\ & 
+ d^L_{ij} (\mu^2) P_L + d^R_{ij} (\mu^2) P_R \Bigr] \frac{1}{p^2-\mu^2 + i \epsilon}, 
\end{align}
where $P_L$ and $P_R$ are the left-handed and right-handed chirality projection operators, respectively, and $c_{ij}^L$, $c_{ij}^R$, $d_{ij}^L$, and $d_{ij}^R$ are the spectral density functions.
By using 
\begin{equation}
\frac{ 1 }{ p^2 - \mu^2 + i \epsilon} \to P \frac{1}{p^2 - \mu^2} - i \pi \delta (p^2 - \mu^2) \quad (\epsilon \to +0), \nonumber
\end{equation}
where $P$ represents the Cauchy principal value,
the dispersive part (the absorptive part) of the two-point functions is derived from the first term (the second term except for $i$).
For clarity, we use the notation $S_{ij}(\cancel{p}; \epsilon)$ to explicitly represent the sign of $i \epsilon$ in Eq.~(\ref{eq: KL_Sij}).

We define the flavor matrix representation for the propagator $S(\cancel{p}; \epsilon)$, whose $(i,j)$ element is $S_{ij}(\cancel{p}; \epsilon)$.
Then, by using the definition of the spectral density functions, 
we can derive the identity~\cite{Grimus:2016hmw}
\begin{equation}
\gamma^0 S(\cancel{p}; \epsilon)^\dagger \gamma^0 = S(\cancel{p}; -\epsilon). \nonumber
\end{equation}
Then, we also have the identity for the inverse propagator $S(\cancel{p}; \epsilon)^{-1}$, 
\begin{equation}
\gamma^0 (S(\cancel{p};\epsilon)^{-1})^\dagger \gamma^0 = S(\cancel{p}; -\epsilon)^{-1}. \nonumber
\end{equation}
The inverse propagator is represented by using the 1PI self-energy $-i \Sigma_{ij}(\cancel{p}; \epsilon)$ as follows:
\begin{equation}
\bigl(S(\cancel{p};\epsilon)^{-1}\bigr)_{ij} = (\cancel{p} - M_i) \delta_{ij} - \Sigma_{ij}(\cancel{p}; \epsilon). \nonumber
\end{equation}
By using this relation, we can obtain the identity for the 1PI self-energy:
\begin{equation}
\label{eq: identity_self-energy}
\gamma^0 \Sigma_{ji}(\cancel{p}; -\epsilon)^\dagger \gamma^0 = \Sigma_{ij}(\cancel{p}; -\epsilon).
\end{equation}

Now, we decompose the 1PI self-energy:
\begin{align}
\label{eq: self-energy_decomposition}
\Sigma_{ij}(\cancel{p}, \epsilon) =&  \cancel{p} \bigl\{ A^L_{ij}(p^2; \epsilon) P_L + A^R_{ij}(p^2; \epsilon) P_R \bigr\} 
\nonumber
\\ & 
+ B^L_{ij}(p^2; \epsilon) P_L + B^R_{ij}(p^2; \epsilon) P_R, 
\end{align}
and define the dispersive part and the absorptive part of $A_L^{ij}$ as 
\begin{equation}
\label{eq: def_Ad_Aa}
A^{L}_{ij}(p^2;\epsilon) = A^{L(d)}_{ij}(p^2) + i A^{L(a)}_{ij}(p^2) 
\end{equation}
and others as well. 
Then, we note that $A^{L}_{ij}(p^2;-\epsilon) = A^{L(d)}_{ij}(p^2) - i A^{L(a)}_{ij}(p^2)$, and so on. 

By using these functions, Eq.~(\ref{eq: identity_self-energy}) leads to 
\begin{align}
\label{eq: identity_for_any_fermions}
A_{ij}^{L(s)}(p^2)^\ast & = A_{ji}^{L(s)}(p^2), \quad 
A_{ij}^{R(s)}(p^2)^\ast = A_{ji}^{R(s)}(p^2), \nonumber
\\
B_{ij}^{L(s)}(p^2)^\ast & = B_{ji}^{R(s)}(p^2), 
\end{align}
where $s= d$ or $a$. The same identity for the dispersive parts is shown in Ref.~\cite{Grimus:2016hmw}. 
Although the dispersive and absorptive parts of each function satisfy the same relation, 
it does not mean $A_{ij}^{L\ast} = A_{ji}^L$ because the complex conjugate changes the sign of $i$ in Eq.~(\ref{eq: def_Ad_Aa}).

Equation~(\ref{eq: identity_for_any_fermions}) holds for both Dirac and Majorana fermions. 
The 1PI self-energy of the Majorana fermions satisfies an additional relation due to their self-conjugate property under the charge conjugate transformation. By using the condition $N_i = N_i^c = C \overline{N}_i^\mathrm{T}$, where $C$ is the charge conjugate matrix, in the two-point function (\ref{eq: 2point_function}), we obtain~\cite{Grimus:2016hmw} 
\begin{equation}
\label{eq: identity_for_Majorana_fermions}
A^{L}_{ij} = A^R_{ij}, \quad 
B^L_{ij} = B^L_{ji}, \quad 
B^R_{ij} = B^R_{ji}, 
\end{equation}
where we omit the common argument $(p^2, \epsilon)$ of the function. 
Finally, by using Eqs.~\eqref{eq: identity_for_any_fermions} and \eqref{eq: identity_for_Majorana_fermions}, the following identities are given for the 1PI self-energy of Majorana fermions: 
\begin{equation}
\label{eq: final_identity_for_Majorana_fermions}
A^{L(s)}_{ij}(p^2)^\ast = A^{R(s)}_{ij}(p^2), \quad 
B^{L(s)}_{ij}(p^2)^\ast = B^R_{ij}(p^2), 
\end{equation}
for $s= d$ and $a$. 
We use these relations to derive the general formula for the $CP$ asymmetry in the decay.

\subsection{$CP$ asymmetry in the decay of $N_i$}

Here, we consider the asymmetry in the decays $N_i \to \psi_\ell V^{(\ast)}$ and its $CP$ conjugate decay $N_i \to \bar{\psi}_\ell V^{\dagger (\ast)}$. 
We assume the following gauge interaction for $N_i$: 
\begin{equation}
\mathcal{L} = - g R_{\ell i} \overline{\psi_\ell} \gamma^\mu P_L N_i V^\dagger_\mu + \mathrm{H.c.} \nonumber
\end{equation}
We consider only the left-handed part of the gauge interaction because it is induced by the mixing with the left-handed neutrinos $\nu_{L}^\ell$ in the lepton doublets. 

First, we consider $\Delta L = 1$ decay, $N_i \to \psi_\ell V^{(\ast)}$. 
The tree level amplitude is given by 
\begin{equation}
i\mathcal{M}_0 = -ig R_{\ell i} \bar{u}_\psi (k)\gamma^\mu P_L u_N(p) j_\mu^\ast, \nonumber
\end{equation}
where $\bar{u}_\psi (k)$ and $u_N(p)$ is the $u$ spinors for $\psi_\ell$ and $N$ with four momenta $k^\mu$ and $p^\mu$, respectively. 
Here, we used the Feynman rules derived according to Ref.~\cite{Denner:1992vza}. 
The form of $j_\mu^\ast$ depends on the process. If we consider the on-shell gauge boson in the final state, $j_\mu$ is the polarization vector of $V_\mu$. If we consider the three-body decays via $V^{\ast}_\mu$, it is the product of the gauge boson propagator and the fermion current. 

To discuss the $CP$ asymmetry in the decay, 
we need to consider the interference term between the tree level and loop level amplitudes~\cite{Kolb:1979qa}. 
As mentioned in Sec.~\ref{Estimation}, we consider only the self-energy contribution as the loop level amplitude. 
Here, we do not limit the discussion to a one-loop level and use the general expression of the self-energy, $-i \Sigma_{ij}$, in the following discussion.

Then, the diagram with the self-energy is given by 
\begin{align}
i\mathcal{M}_1 = & -\frac{ igR_{\ell j} }{ M_i^2 - M_j^2 } \bar{u}_\psi(k) \gamma^\mu P_L (\cancel{p} + M_j) \Sigma_{ij} (\cancel{p}) u_N(p)j_\mu^\ast \nonumber \\
= & -\frac{ igR_{\ell j} }{ M_i^2 - M_j^2 } \bar{u}_\psi(k) \gamma^\mu C_{ij}(M_i^2) P_L u_N(p)j_\mu^\ast,\nonumber
\end{align}
where the sum over $j$ $(\neq i)$ is implicitly taken, and the function $C_{ij}(p^2)$ is defined by
\begin{align}
C_{ij}(p^2) = & M_i^2 A^L_{ij}(p^2) + M_i M_j A^R_{ij}(p^2) 
\nonumber
\\ & 
+ M_i B^R_{ij}(p^2) + M_j B^L_{ij}(p^2), \nonumber
\end{align}
where $A^L_{ij}$, $A^R_{ij}$, $B^L_{ij}$ and $B^R_{ij}$ are defined in the self-energy decomposition in Eq.~(\ref{eq: self-energy_decomposition}).
The interference terms of $\mathcal{M}_0$ and $\mathcal{M}_1$ are given by
\begin{align}
& \overline{\mathcal{M}_0 \mathcal{M}_1^\ast} = \frac{ 1 }{ 2 } \frac{ g^2 R_{\ell i} R_{\ell j}^\ast }{ M_i^2 - M_j^2 } \biggl( \sum_\mathrm{spin} j_\mu^\ast j_\nu \biggr) C_{ij}^\ast \mathrm{tr}[\cancel{k}\gamma^\mu \cancel{p} \gamma^\nu P_L], \nonumber \\[5pt]
& \overline{\mathcal{M}_0^\ast \mathcal{M}_1} = \frac{ 1 }{ 2 } \frac{ g^2 R_{\ell i}^\ast R_{\ell j} }{ M_i^2 - M_j^2 } \biggl( \sum_\mathrm{spin} j_\mu^\ast j_\nu \biggr) C_{ij} \mathrm{tr}[\cancel{p}\gamma^\nu \cancel{k} \gamma^\mu P_L], \nonumber 
\end{align}
where we take the spin average for $N_i$ and the spin sum for particles in the final state.

Next, we consider the $CP$ conjugate decays $N_i \to \bar{\psi}_\ell V^{\dagger (\ast)}$. 
The tree-level amplitude $\tilde{\mathcal{M}}_0$ and the loop level amplitude $\tilde{\mathcal{M}}_1$ are given by
\begin{align}
& \tilde{\mathcal{M}}_0 = -ig R_{\ell i}^\ast \bar{v}_\psi (k) \gamma^\mu P_R u_N(p) j_\mu, \nonumber \\[5pt]
& \tilde{\mathcal{M}}_1 = - \frac{ i g R_{\ell j}^\ast }{ M_i^2 - M_j^2 } \bar{v}_\psi (k) \gamma^\mu D_{ij}(M_i^2) P_R u_N(p), \nonumber
\end{align}
where $v_\psi$ is the $v$ spinor for $\psi_\ell$, and the function $D_{ij}(p^2)$ is defined as 
\begin{align}
D_{ij}(p^2) = & M_i M_j A^L_{ij}(p^2) + M_i^2 A^R_{ij}(p^2) 
\nonumber
\\ & 
+ M_i B^L_{ij}(p^2) + M_j B^R_{ij}(p^2).\nonumber
\end{align}
Then, the interference terms are given by 
\begin{align}
& \overline{\tilde{\mathcal{M}}_0 \tilde{\mathcal{M}}_1^\ast } = \frac{ 1 }{ 2 } \frac{ g^2 R_{\ell i}^\ast R_{\ell j} }{ M_i^2 - M_j^2 } \biggl( \sum_\mathrm{spin} j_\mu^\ast j_\nu \biggr) D_{ij}^\ast \mathrm{tr}[\cancel{k}\gamma^\nu \cancel{p} \gamma^\mu P_R], \nonumber \\[5pt]
& \overline{\tilde{\mathcal{M}}_0^\ast \tilde{\mathcal{M}}_1 } = \frac{ 1 }{ 2 } \frac{ g^2 R_{\ell i} R_{\ell j}^\ast }{ M_i^2 - M_j^2 } \biggl( \sum_\mathrm{spin} j_\mu^\ast j_\nu \biggr) D_{ij} \mathrm{tr}[\cancel{p}\gamma^\mu \cancel{k} \gamma^\nu P_R]. \nonumber
\end{align}

We define the quantity $\Delta \equiv \ \overline{\mathcal{M}_0 \mathcal{M}_1^\ast} + \overline{\mathcal{M}_0^\ast \mathcal{M}_1} - \overline{\tilde{\mathcal{M}}_0 \tilde{\mathcal{M}}_1^\ast } - \overline{\tilde{\mathcal{M}}_0^\ast \tilde{\mathcal{M}}_1 }$. 
Using the above equations, $\Delta$ is given by
\begin{align}
\Delta = \frac{g^2}{2} \frac{s_{\mu\nu}a^{\mu\nu} + t_{\mu\nu} b^{\mu\nu} }{ M_i^2 - M_j^2 }
\Bigl\{ & R_{\ell i} R_{\ell j}^\ast (C_{ij}^\ast - D_{ij}) 
\nonumber
\\ & 
+ R_{\ell i}^\ast R_{\ell j} (C_{ij} - D_{ij}^\ast ) \Bigr\}, \nonumber
\end{align}
where 
\begin{align}
& s_{\mu \nu} = \frac{ 1 }{2 } \biggl( \sum_\mathrm{spin} j_\mu^\ast j_\nu + j_\nu^\ast j_\mu \biggr), \quad t_{\mu\nu} = \frac{1 }{ 2 } \biggl( \sum_\mathrm{spin} j_\mu^\ast j_\nu - j_\nu^\ast j_\mu \biggr), \nonumber \\
& a^{\mu\nu} = \frac{ 1 }{ 2 } \mathrm{tr}[\cancel{p}\gamma^\mu \cancel{k} \gamma^\nu], \quad b^{\mu\nu} = \frac{ 1 }{ 2 } \mathrm{tr}[\cancel{p}\gamma^\mu \cancel{k}\gamma^\nu \gamma_5]. \nonumber
\end{align}
We note that the process-dependent part is factorized by $s_{\mu\nu} a^{\mu\nu} + t_{\mu\nu} b^{\mu\nu}$ in $\Delta$. 

By using $\Delta$, the $CP$ asymmetry is given by
\begin{align}
\varepsilon_\ell^{(i)} & \equiv \frac{ \Gamma (N_i \to \psi_\ell V) - \Gamma (N_i \to \bar{\psi}_\ell V^\dagger) }{ \Gamma (N_i \to \psi_\ell V) +\Gamma (N_i \to \bar{\psi}_\ell V^\dagger) } \nonumber
\\ &
\simeq  \frac{ \int d\Pi \, \Delta }{ \int d\Pi \, (\overline{|\mathcal{M}_0|}^2 + \overline{|\mathcal{M}_0|}^2) }, \nonumber
\end{align}
where $\int \mathrm{d} \Pi$ is the phase space integral. 
The tree-level squared amplitudes are given by
\begin{equation}
\overline{|\mathcal{M}_0|}^2 = \overline{|\mathcal{M}_0|}^2 = \frac{g^2}{2} |R_{\ell i}|^2 (s_{\mu\nu} a^{\mu\nu} + t_{\mu\nu} b^{\mu\nu} ). \nonumber
\end{equation}
Therefore, the process-dependent part, $\int d \Pi\,  (s_{\mu\nu} a^{\mu\nu} + t_{\mu\nu} b^{\mu\nu})$ is completely canceled in the denominator and numerator.
$\varepsilon_\ell^{(i)}$ is given by
\begin{align}
\varepsilon_\ell^{(i)} = & \frac{ 1 }{ 2 (M_i^2 - M_j^2) |R_{\ell i}|^2 } 
\Bigl\{ R_{\ell i} R_{\ell j}^\ast \Bigl(C_{ij}^\ast (M_i^2) - D_{ij} (M_i^2) \Bigr) 
\nonumber
\\ & 
+ R_{\ell i}^\ast R_{\ell j} \Bigl(C_{ij}(M_i^2) - D_{ij}^\ast (M_i^2) \Bigr) \Bigr\}. \nonumber
\end{align}
Thus, $\varepsilon_\ell^{(i)}$ generated by the self-energy is simply determined by the Yukawa coupling $y_{\ell i}^{}$, the mass $M_i$, and the self-energy at $p^2 = M_i^2$. 

Next, we evaluate $C_{ij}^\ast - D_{ij}$. 
By using their definitions, 
\begin{align}
C_{ij}^\ast - D_{ij} = & M_i^2 (A_{ij}^{L\ast} - A_{ij}^R) + M_i M_j (A_{ij}^{R\ast} - A_{ij}^{L}) 
\nonumber
\\ & 
+ M_i (B_{ij}^{R\ast} - B_{ij}^{L}) + M_j (B_{ij}^{L\ast} - B_{ij}^{R\ast} ). \nonumber
\end{align}
The identity for the 1PI self-energy of Majorana femrions in Eq.~\eqref{eq: final_identity_for_Majorana_fermions} leads to 
\begin{align}
C_{ij}^\ast - D_{ij} = & - 2i \Bigl\{ 
    M_i \Bigl( M_i A_{ij}^{R(a)} + B^{L(a)}_{ij} \Bigr) 
    \nonumber
\\ & 
    + M_j \Bigl( M_i A_{ij}^{R(a)} + B^{L(a)}_{ij} \Bigr)^\ast\Bigr\}. \nonumber
\end{align}
We note that contributions from the dispersive parts are canceled in $C_{ij}^\ast - D_{ij}$. 
Thus, if the self-energy $\Sigma_{ij}(\cancel{p})$ does not include the absorptive part at $p^2 = M_i^2$, then the $CP$ asymmetry in the decay is zero even if the Yukawa coupling $y_{\ell i}^{}$ includes $CP$-violating phases. 
This is a well-known consequence of the $CPT$ invariance and the unitarity~\cite{Kolb:1979qa}. 
Thus, we obtain
\begin{align}
\label{eq: epsilon_general}
\varepsilon_\ell^{(i)} = & \frac{ -i }{ (M_i^2 - M_j^2) |R_{\ell i}|^2 } 
\nonumber
\\ & 
\Bigl\{ 
\Bigl( M_i R_{\ell i} R_{\ell j}^\ast - M_j R_{\ell i}^\ast R_{\ell j} \Bigr) \Bigl( M_i A_{ij}^{R(a)} + B_{ij}^{L(a)} \Bigr) \nonumber \\
& + \Bigl( M_j R_{\ell i} R_{\ell j}^\ast - M_i R_{\ell i}^\ast R_{\ell j} \Bigr) \Bigl( M_i A_{ij}^{R(a)} + B_{ij}^{L(a)} \Bigr)^\ast 
\Bigr\}.
\end{align}

Equation~(\ref{eq: epsilon_general}) is a general expression for $\varepsilon_\ell^{(i)}$ without assumption. 
Here, we impose a practical assumption on the self-energy $-i \Sigma_{ij}$, i.e., we consider only the leading contribution of the couplings and neglect lepton masses. 
We use the spurion technique to derive simple formulas for $A_{ij}^{R(a)}$ and $B_{ij}^{L(a)}$ with this simplification. 

The Lagrangian relevant to $N_i$ is 
\begin{align}
\label{eq: N_Lagrangian}
\mathcal{L} \simeq & -\frac{ 1 }{ 2 } (M_N)_{ij} \overline{N^i_R} (N^j_R)^c - \frac{g_W^{}}{\sqrt{2}} R_{ai} \bar{\ell}_a \gamma^\mu (N^i_R)^c W_{\mu}^- \nonumber \\[5pt]
& - \frac{g_Z^{}}{2} R_{ai} \overline{\nu_a} \gamma^\mu (N^i_R)^c Z_{\mu} - \frac{vy_{ai}}{ \sqrt{2} } h \overline{\nu_a} N^i_R + \mathrm{H.c.},
\end{align}
in addition to their kinetic terms, where we used the right-handed fields, $N^i_R = P_R N_i$. 
Let us consider a unitary rotation $N^i_R \to X_{ij} N^j_R$, where $X$ is a unitary matrix.
Then, to make the Lagrangian (\ref{eq: N_Lagrangian}) invariant with this rotation, the spurion fields $M_{ij}$, $R_{ai}$, and $y_{ai}$ has to transform as 
\begin{align}
& (M_N)_{ij} \to X^\ast_{ik} X_{jm}^\ast (M_N)_{km}, 
\nonumber
\\ & 
R_{ai} \to X_{ik} R_{a k}, \quad y_{ai}^{} \to X_{ik}^\ast y_{ak}. \nonumber
\end{align}
This transformation is consistent with $R = (v/\sqrt{2}) y M_N^{-1}$.
In Eq.~(\ref{eq: N_Lagrangian}), we omit the coupling between $\phi$ and $N_i$ irrelevant to the current discussion. 
If we include it, then it must transform in the same way as $M_N$. 
In addition to this symmetry, we impose the lepton number conservation. Then, $R$ and $y$ carry the lepton number $+1$, while $M_N$ has no lepton number because we do not assign the lepton number to $N_i$. 

First, we consider the transformation of $A^R_{ij}$. 
Since it corresponds to the 1PI two-point function $\bigl< T (N_R^i)^c \overline{(N_R^j)^c} \bigr>_\mathrm{1PI}$, 
it transforms as 
\begin{equation}
A_{ij}^R \to X_{ik}^\ast X_{jm} A_{km}^R. \nonumber
\end{equation}
Thus, at the leading order of the coupling, its flavor structure is given by
\begin{equation}
    (R^\dagger R)_{ij}, \quad 
    (R^\mathrm{T} y)_{ij}, \quad
    (R^\mathrm{T}RM)_{ij}, \nonumber
\end{equation}
because we neglect the lepton masses. 
However, the second and third terms carry the lepton number $+2$ and are thus prohibited. 
Therefore, $A_{ij}^{R} \propto (R^\dagger R)_{ij}$.

In the same way, we can show that $B_{ij}^L$ transforms as $B_{ij}^L \to X_{ik}^\ast X_{jm}^\ast B_{km}^L$, and the flavor structure is given by
\begin{equation}
\begin{array}{lll}
(R^\dagger R^\ast)_{ij}, & (R^\dagger R M)_{ij}, & (M R^\mathrm{T} R^\ast)_{ij}, \\
(y^\dagger y^\ast)_{ij}, & (y^\dagger y M)_{ij}, & (M y^\mathrm{T} y^\ast)_{ij}, \\
\end{array} \nonumber 
\end{equation}
at the leading order. $R^\dagger R^\ast$ and $y^\dagger y^\ast$ violate the lepton number and are prohibited. 
Although the others are allowed by symmetry, we can show that it is impossible to make such Feynman diagrams for the 1PI self-energy as follows. Two vertices of $R$ ($y$) and $R^\dagger$ ($y^\dagger$) must each be connected to the external lines. Thus, the corresponding Feynman diagrams do not include the internal lines of $N_i$ because there is no other coupling to generate $N_i$. Thus, $M_N$ cannot be included in the amputated diagrams.

Consequently, at the leading order of the coupling, we obtain 
\begin{equation}
A_{ij}^{R(a)} = (R^\dagger R)_{ij} A(p^2), \quad 
B_{ij}^{L(a)} = 0, \nonumber
\end{equation}
where $A(p^2)$ is a scalar function that does not include the flavor indices. Using these, the leading $CP$ asymmetry from the self-energy diagrams is given by
\begin{align}
\varepsilon_\ell^{(i)} = & \frac{ -i M_i A(M_i^2) }{ (M_i^2 - M_j^2) |R_{\ell i}|^2 } 
\nonumber
\\ & 
\Bigl\{
M_i \Bigl(R_{\ell i} R_{\ell j}^\ast(R^\dagger R)_{ij} - R_{\ell i}^\ast R_{\ell j}(R^\dagger R)_{ji} \Bigr) \nonumber \\
& + M_j \Bigl(R_{\ell i} R_{\ell j}^\ast (R^\dagger R)_{ji} - R_{\ell i}^\ast R_{\ell j} (R^\dagger R)_{ij}\Bigr) \Bigr\}. \nonumber
\end{align}
The scalar function $A(M_i^2)$ can be represented by the total decay rate $\Gamma_i$ using the optical theorem.\footnote{Obviously, the total decay rate is also of the quadratic order of the coupling.} By considering the diagonal two-point function at $p^2 = M_i^2$, the theorem leads to
\begin{equation}
\Gamma_i = - 2 (R^\dagger R)_{ii}  M_i A(M_i^2) . \nonumber
\end{equation}

Thus, we obtain the final form of $\varepsilon_\ell^{(i)}$: 
\begin{align}
\varepsilon_\ell^{(i)} = & \frac{ \Gamma_i }{ 2 |R_{\ell i}|^2 (R^\dagger R)_{ii}} 
\nonumber
\\ & 
\sum_{j\neq i} \frac{ 1 }{ M_i^2 - M_j^2 } \Bigl\{ 
i M_i \Bigl(R_{\ell i} R_{\ell j}^\ast(R^\dagger R)_{ij} - R_{\ell i}^\ast R_{\ell j}(R^\dagger R)_{ji} \Bigr) \nonumber \\
& + i M_j \Bigl(R_{\ell i} R_{\ell j}^\ast (R^\dagger R)_{ji} - R_{\ell i}^\ast R_{\ell j} (R^\dagger R)_{ij}\Bigr) \Bigr\}, \nonumber
\end{align}
where we show the sum over $j$ explicitly.

We note that $\varepsilon_\ell^{(i)}$ is the $CP$ asymmetry in one decay process, which satisfies 
\begin{equation}
(\Gamma_\sigma^\ell + \Gamma_{\bar{\sigma}}^\ell) \varepsilon_{\ell}^{(i)} = (\Gamma_\sigma^\ell - \Gamma_{\bar{\sigma}}^\ell), \nonumber
\end{equation}
where $\Gamma^{\ell}_\sigma$ ($\Gamma^{\ell}_{\bar{\sigma}}$) represents the rate of the considered decay (that of its $CP$ conjugate decay), $\sigma$ is a label for the process, and $\ell$ represents the process that produces (annihilates) one lepton number with the lepton flavor $\ell$.

To solve the density matrix equation in Eq.~\eqref{NBLeqnres}, we need a $CP$ asymmetry for the total decay width satisfying 
\begin{equation}
\Gamma_i \varepsilon_{\ell \ell}^{(i)} = \sum_{\sigma} (\Gamma_\sigma^\ell - \Gamma_{\bar{\sigma}}^{\ell} ).\nonumber
\end{equation}
Then, $\varepsilon_{\ell \ell}^{(i)}$ is given by
\begin{align}
\varepsilon_{\ell \ell}^{(i)} = \frac{\sum_{\sigma} (\Gamma_\sigma^\ell - \Gamma_{\bar{\sigma}}^{\ell} ) }{ \sum_\ell \sum_\sigma (\Gamma_\sigma^\ell + \Gamma_{\bar{\sigma}}^\ell )}
= \frac{\varepsilon_{\ell}^{(i)} \sum_{\sigma} (\Gamma_\sigma^\ell + \Gamma_{\bar{\sigma}}^{\ell} ) }{ \sum_\ell \sum_\sigma (\Gamma_\sigma^\ell + \Gamma_{\bar{\sigma}}^\ell )}, \nonumber
\end{align}
where we used the fact that $\varepsilon^{(i)}_\ell$ is process independent. 
Since we evaluate $\varepsilon_{\ell \ell}^{(i)}$ at the leading order, we can use the tree-level expression for $\Gamma_\sigma^\ell$ and $\Gamma_{\bar{\sigma}}^\ell$: 
\begin{equation}
    \Gamma_\sigma^\ell = \Gamma_{\bar{\sigma}}^\ell = |R_{\ell i}|^2 C_\sigma, \nonumber
\end{equation}
where $C_\sigma$ is a coefficient not including the lepton flavor indices. 
Using this, we get
\begin{equation}
\label{eq: total_and_partial_CPasym}
    \varepsilon_{\ell \ell}^{(i)} = \varepsilon_{\ell}^{(i)} \frac{ |R_{\ell i}|^2 }{ (R^\dagger R)_{ii}}. 
\end{equation}

Finally, we obtain the following final form of the $CP$ asymmetry in the total decay
\begin{align}
\varepsilon_{\ell\ell}^{(i)} = & \frac{ \Gamma_i }{ 2 M_i \{(R^\dagger R)_{ii}\}^2 } \frac{ (M_i^2 - M_j^2) M_i^2}{ (M_i^2-M_j^2)^2 + M_i^4 \Gamma_j^2/M_j^2}
\nonumber \\
& \sum_{j\neq i} 
\Bigl\{
i \Bigl(R_{\ell i} R_{\ell j}^\ast (R^\dagger R)_{ji} - R_{\ell i}^\ast R_{\ell j} (R^\dagger R)_{ij}\Bigr) \frac{M_j}{M_i}\nonumber \\
& + i \Bigl(R_{\ell i} R_{\ell j}^\ast(R^\dagger R)_{ij} - R_{\ell i}^\ast R_{\ell j}(R^\dagger R)_{ji} \Bigr) \Bigr\} ,
\end{align}
where we replaced $1/(M_i^2 -M_j^2)$ with 
\begin{equation}
\frac{1}{M_i^2 -M_j^2} \to \frac{M_i^2 - M_j^2}{ (M_i^2 - M_j^2)^2 + M_i^4 \Gamma_j^2/M_j^2 }. \nonumber
\end{equation}
This is the diagonal term for the complex matrix $\varepsilon_{\alpha \beta}^{(i)}$. The nondiagonal term is given by replacing the indices appropriately.

Here, we consider only processes mediated by a single gauge boson. However, an extension to processes mediated by multiple gauge bosons is straightforward, and we can use the same expression for $\varepsilon_{\alpha \beta}^{(i)}$.

It is also straightforward to extend the above discussion to the process including the Higgs boson, $N_i \to \nu_\ell h^{(\ast)}$. In this case, $R_{\ell i}$ in the above discussion must be replaced by $y_{\ell i}$. Thus, if we consider this process along with the gauge boson-mediated processes, we have to treat it separately because we assume that all the decay rates are proportional to $|R_{\ell i}|^2$ at tree level in deriving Eq. \eqref{eq: total_and_partial_CPasym}, which is not valid in this case. However, as discussed in Sec.~\ref{Estimation}, when we consider very small mass differences, $|M_i - M_j| \ll M_i, M_j$, the difference between the $CP$ asymmetries using $y_{\ell i}$ and $R_{\ell i}$ is sufficiently small, and we can use the same expression for $\varepsilon^{(i)}_{\alpha \beta}$ as that before the EWSB throughout all eras of the early Universe.

\bibliography{ref}
\end{document}